\documentclass[aps,prx,twocolumn,showpacs,superscriptaddress,floatfix,10pt,amsmath,amssymb,footinbib]{revtex4-2}

\usepackage{graphicx}
\usepackage{bm,bbm}
\usepackage[urlcolor=blue]{hyperref}
\hypersetup{colorlinks=true,allcolors=blue}

\usepackage{dcolumn}
\usepackage{amsmath,amscd,amsbsy,amssymb,latexsym,url,amsthm}

\begin{document}

\title{On the emergence of memory 
  in equilibrium versus non-equilibrium systems}

\author{Xizhu Zhao}
\affiliation{Mathematical bioPhysics Group, Max Planck Institute for Multidisciplinary Sciences, Am Fa\ss berg 11, 37077 G\"ottingen}
\affiliation{Max Planck School Matter to Life, Jahnstraße 29, D-69120 Heidelberg, Germany}
\author{David Hartich}
\author{Alja\v{z} Godec}
\email{agodec@mpinat.mpg.de}
\affiliation{Mathematical bioPhysics Group, Max Planck Institute for
  Multidisciplinary Sciences, Am Fa\ss berg 11, 37077 G\"ottingen}
\date{\today}

\begin{abstract}
Experiments often probe observables that correspond to low-dimensional
projections of high-dimensional dynamics. In such situations distinct microscopic
configurations become lumped into the same observable
state. It is well known that correlations between the observable and
the hidden degrees of freedom give rise to memory effects. However,
how and under which conditions these correlations emerge remains
poorly understood. Here we shed light on two fundamentally different
scenarios of the emergence of memory in minimal stationary systems,
where observed and hidden degrees of freedom evolve either
cooperatively or are 
coupled by a hidden non-equilibrium current. In the
reversible setting strongest memory manifests when the time-scales of
hidden and observed dynamics 
overlap, whereas, strikingly, in the driven
setting maximal memory emerges under a clear
time-scale separation. Our results hint at the possibility of 
fundamental differences in the way memory emerges in
equilibrium versus driven systems \textcolor{black}{that may be
  utilized as a ``diagnostic'' of the underlying hidden transport mechanism}.
\end{abstract}

\maketitle
Observables coupled to hidden degrees of freedom that 
do not relax sufficiently fast \cite{lapolla2019manifestations} or
selected reaction coordinates that do not locally equilibrate in 
meso-states \cite{PhysRevX.11.041047} generically display memory. In fact,
this holds for most high-dimensional dynamics probed on a coarse-grained level
\cite{Wolynes,Xie_2005,JCS_2008,sangha_proteins_2009,avdoshenko_theoretical_2017,
	Dima_2013,Dima_2019,Meyer_2020,woodside,Netz,Krueger,Solano}. Tremendous
progress has been made over the years in describing and understanding
kinetic aspects of non-Markovian dynamics         
\cite{Haken,Haenggi_1,Haenggi_2,Sancho,Grigolini,Goychuk,Igor,Igor_2,Igor_3,Metz_1,Metz_2,Eli_1,Eli_2,Eli_3,Eli_Igor,Satya2,Satya_c1,Satya_critical,lapolla2019manifestations,Lapolla_2020,Haenggi_RMP,Netz,Baiesi_GLE,netz2023derivation,netz2023multipoint}. More 
recently, coarse-grained, partially observed dynamics have become of
great interest from the point of view of thermodynamic inference 
\cite{Baiesi_TUR,PhysRevLett.108.220601,Massi,van2022thermodynamic,PhysRevX.11.041047,Polettini,Need,Snippets,doi:10.1021/acs.jpclett.2c03244,andrieux2012bounding,PhysRevE.78.011107,rahav2007fluctuation,PhysRevLett.125.110601,PhysRevLett.105.150607,PhysRevE.85.031129,baiesi2023effective,netz2023multipoint}. Namely,
while efficient methods exist to detect \cite{berezhkovskii_single-molecule_2018,Lindner,PhysRevResearch.3.L022018} and quantify \cite{PhysRevResearch.3.L022018}
the existence of memory,  
it \textcolor{black}{conversely} turns out to be quite
challenging to quantify \cite{PhysRevX.11.041047,Polettini,Snippets,Viol} or even
infer \cite{God_Mak,PhysRevLett.105.150607,PhysRevE.85.031129,netz2023multipoint}
irreversibility from lower-dimensional, projected
dynamics.  \textcolor{black}{Thus, understanding potential differences in the emergence of memory in equilibrium and non-equilibrium systems is a difficult task.}

Considering in particular ergodic dynamics in the sense that the
\textcolor{black}{probability distribution to be found in a given
  microscopic state at long times} relaxes to a \textcolor{black}{unique}
stationary, equilibrium or non-equilibrium, steady state from any
initial condition, the extent of memory is  
necessarily finite and is more prominent if the hidden degrees of
freedom are slow
\cite{lapolla2019manifestations,TUR_David,PhysRevX.11.041047}. Yet,
even in this ``well behaved'', thermodynamically consistent \cite{seifert2012stochastic} setting quite little is known about the 
possible ways in which the dynamics of observables can become
correlated with that of hidden degrees of
freedom on different time scales. A particularly intriguing question
is whether there are any characteristic differences between how
memory emerges in reversible versus irreversible, driven systems when
observed dynamics is much faster than the hidden one.   
	\begin{figure}[h]
		\includegraphics[width=0.49\textwidth]{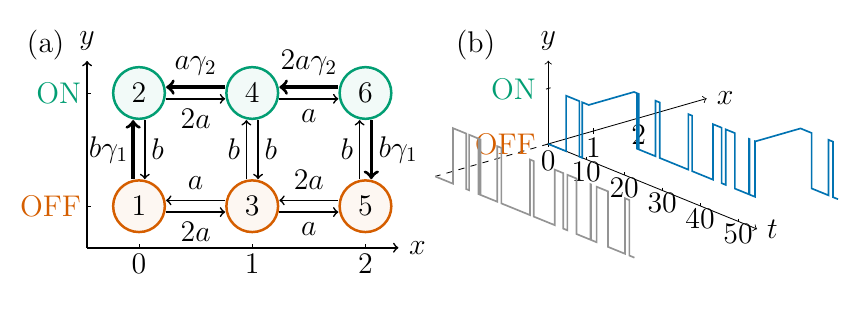}
		\caption{(a) Schematics of the full Markov network for
                  the respective models. The upper green
                  states are lumped to the observed ``ON'' state and the lower 
                  orange ones to the observed ``OFF'' state. In order to separate the time
                  scale of horizontal and vertical dynamics, we choose
                  the rates (in arbitrary units)
                  $a=0.5$ and $b=25$ to make vertical process always
                  much faster than horizontal one. In the driven model
                  (see Fig.~\ref{fig2}) we set $\gamma_1>1$ and
                  $\gamma_2=1$, whereas in the reversible
                  ``allosteric'' model (Fig.~\ref{fig3}) we chose
                  $\gamma_1=\gamma_2\equiv\gamma\ge 1$. A trajectory of the full
                  dynamics (blue) and
                  its corresponding projection (gray) are shown in
                  panel (b).}
                \label{fig1}
	\end{figure}

Addressing this problem in full generality is a daunting task. Here we
focus on the minimal ``cooperative'' setting \cite{MWC,Tu,Cluzel,marzen2013statistical},
where the microscopic dynamics is a Markov process on a planar network and we
observe only the vertical 
coordinate whereas the 
horizontal 
transitions are hidden (see Fig.~\ref{fig1}). \textcolor{black}{This
particular setting is important for understanding  ``active secondary
transport''---transporter proteins exploiting the energy stored in
transmembrane gradient of one type of molecules to transport another
type against their
gradients~\cite{Tolya_1,Tolya_2,Tolya_3}. Moreover, it is also
relevant as the physical basis of
the 
sensitivity of the flagellar motor in \emph{E.~coli}
in sensing concentrations of its regulator
\cite{MWC,Tu,Cluzel,hartich2015nonequilibrium}, where two
fundamentally different explanations were proposed to explain the
``ultrasensitivity'' of the motor's response, an
equilibrium allosteric 
and a dissipative non-equilibrium model  \cite{Tu,hartich2015nonequilibrium,PhysRevX.10.011066}.}

We are not
interested in the biophysical implications of the model. Instead, we
use the above two distinct settings merely as a minimal model of non-Markovian
2-state dynamics, where we can ``turn on'' memory in a controlled
manner via an equilibrium versus
non-equilibrium mechanism. We consider the scenario,
where the observed dynamics is faster than the hidden (see
Fig.~\ref{fig1}a) but there is \emph{not} necessarily a large time-scale
separation present.  Interestingly, in the driven setting
we only change the fastest time scales, whereas in the reversible
cooperative model we alter all time scales.
In both cases we find a maximal capacity
for memory, i.e.\ the maximal magnitude of memory saturates at a
finite coupling and non-equilibrium driving,
respectively. Interestingly, in the reversible setting the memory
manifests strongest when the time-scale
separation between hidden and observed transition becomes partially
lifted and the hidden and observable time-scales overlap, whereas in the driven setting maximal memory
occurs in the presence of a time-scale separation. Our
results provide deeper insight into the emergence of memory in the 
distinct situations when observed and hidden dynamics either evolve
cooperatively or become coupled by a hidden non-equilibrium current. 

\emph{Setup.---}We consider a 6-state continuous-time Markov process
(see Fig.~\ref{fig1}a)
with generator $L$, whose elements $L_{n,m}$ are transition rates between
states $m\to n$ given by
\begin{widetext}
			\begin{equation} 
				L=\displaystyle{ 
				\begin{pmatrix}
                                \ 
					-b\gamma_1-2a & b & a & 0 & 0 & 0 \\
					b\gamma_1 & -b-2a & 0 & a \gamma_2 & 0 & 0 \\
					2a & 0 & -b-2a & b & 2a & 0 \\
					0 & 2a & b & -b -a -a\gamma_2 & 0 & 2a\gamma_2 \\
					0 & 0 & a & 0 & -b-2a & b\gamma_1 \\
					0 & 0 & 0 & a & b & -b\gamma_1 - 2a \gamma_2
				\end{pmatrix}}, 
			\end{equation}
\end{widetext}
where we choose $a=1/2$ and $b=25$ such that in the ``baseline'' model
horizontal transitions are slower than vertical ones, i.e.\ the hidden
dynamics relax slower than the observable ON$\rightleftarrows$OFF
transitions. We adopt the Dirac bra-ket notation and denote the transition probability from microscopic state $i$ to
microscopic state $j$ in time $t$ as $G(j,t|i)\equiv\langle j|\mathrm{e}^{Lt}|i\rangle$
and the stationary probability of
state $j$ as $P_s(j)=\lim_{t\to\infty}G(j,t|i)$.  In the baseline model with $\gamma_1=\gamma_2 = 1$ the
dynamics is reversible and the ON/OFF states are equi-probable in the
steady state. The parameters $\gamma_1$
and $\gamma_2$ are \emph{acceleration factors} when they are larger
than 1. In the cooperative allosteric regime with $\gamma_1 = \gamma_2\equiv\gamma> 1$ the dynamics 
obeys
detailed balance, i.e.\ $P_s(i) L_{j,i}= P_s(j) L_{j,i},\,\forall i,j$. Conversely, 
in the driven model we set $\gamma_1>1, \gamma_2=1$; here detailed balance is violated, i.e.\ $\exists i,j$ for which $P_s(i) L_{j,i} \neq P_s(j) L_{j,i}$. 
 (that is, the model is irreversible yet thermodynamically
consistent \cite{seifert2012stochastic}). 

It turns out that $L$ defined this way is diagonalizable, i.e.\ we can
find a bi-orthonormal basis $\{\langle\psi_k^L |,|\psi_k^R\rangle\}$
of left $\langle\psi_k^L |$ and right $|\psi_k^R\rangle$ eigenvectors
with eigenvalue $-\lambda_k$ and $k=0,\ldots,5$ and
$\langle\psi_k^L|\psi_l^R\rangle = \delta_{kl}$. Thus, we have
$L=\sum_k -\lambda_k |\psi_k^R\rangle\langle\psi_k^L|$ and in turn we
can expand $G(j,t|i)=\sum_{k=0}^5\langle
j|\psi_k^R\rangle\langle\psi_k^L|i\rangle e^{-\lambda_k t}$. In this
notation the steady-state probability of state $i$ is \textcolor{black}{given by}
$P_s(i)=\langle i |\psi_0^R\rangle$.

The eigenvalues of the baseline model are $\lambda_0=0, \lambda_1=2a,
\lambda_2=4a, \lambda_3=2b, \lambda_4=2(a+b), \lambda_5=2(2a+b)$. We
can also determine the eigenspectrum analytically for the driven
model, which has eigenvalues $\lambda^{\gamma_1}_0=0, \lambda^{\gamma_1}_1=2a, \lambda^{\gamma_1}_2=4a,
\lambda^{\gamma_1}_3=\frac{1}{2}[4a+(\gamma_1+3)b-\sqrt{16a^2+(\gamma_1
  -1)^2b^2}], \lambda^{\gamma_1}_4=2a+(\gamma_1+1)b,
\lambda^{\gamma_1}_5=\frac{1}{2}[4a+(\gamma_1+3)b+\sqrt{16a^2+(\gamma_1
  -1)^2b^2}]$. The reversible allosteric model ($\gamma > 1$) cannot be diagonalized analytically and we therefore provide numerical results instead.

We assume that the full system is prepared in a steady state $P_s(j)$
and only vertical ON$\rightleftarrows$OFF transitions are
observed with observable sets ${\rm ON}=\{2,4,6\}$ and ${\rm OFF}=\{1,3,5\}$. 
\textcolor{black}{
We determine the non-Markovian transition probability of the observed process $\hat{k}_t$,  $Q_{P_s}(\hat{n},t|\hat{m})$ with $\hat{m},\hat{n}\in\{{\rm ON},{\rm OFF}\}$ as \cite{lapolla2019manifestations} 
\begin{equation}
	Q_{P_s}(\hat{n},t|\hat{m})\equiv\frac{\sum_{j=1}^6\mathbbm{1}_{\hat{n}}[j]\sum_{i=1}^6\mathbbm{1}_{\hat{m}}[i]G(j,t|i)P_{s}(i)}{\sum_{i=1}^6\mathbbm{1}_{\hat{m}}[i]P_{s}(i)},
	\label{pTr}  
\end{equation}
where $\mathbbm{1}_{\Omega}$ is the indicator function of the set $\Omega$. }
The non-Markovian transition
probability between two fixed observed states $\hat{m}\to\hat{n}$
as well as the observable return probability $\hat{m}\to\hat{m}$
depend on the preparation of the full system \cite{lapolla2021ubiquitous}. Moreover,
in spite of the \emph{full} system being prepared in the
stationary state $P_s$, by specifying the initial observed state (here either
``ON'' or ``OFF'') we ``quench'' the full system out of the steady state
by conditioning on the state of the observable
\cite{lapolla2019manifestations,lapolla2021ubiquitous}. Without loss
of generality we will focus on the scenario where the observable is
initially in the ON
state, i.e.\ $\hat{k}_0={\rm ON}$.

To quantify the magnitude 
of memory
in the projected dynamics, we follow \cite{PhysRevResearch.3.L022018}
and construct the auxiliary Chapman-Kolmogorov (CK) transition density
\begin{equation}
  Q^{\rm CK}(\hat{n},t_1+t_2|\hat{m}) \equiv
  \sum_{\hat{k}}
  Q_{P_s}(\hat{n},t_2|\hat{k})Q_{P_s}(\hat{k},t_1|\hat{m}).
  \label{Gck}
\end{equation}
Note
that for a non-Markovian process $Q^{\rm CK}$ depends on both $t_1$
and $t_2$. The Chapman-Kolmogorov construction $Q^{\rm CK}(\hat{n},t_1+t_2|\hat{m}) $ corresponds
to a fictitious dynamics where we \emph{force} at time $t_1$ all hidden degrees of freedom to
their stationary distribution and thereby erase all memory of their
initial condition. 
When the observed ON$\rightleftarrows$OFF dynamics is Markovian we
have $Q^{\rm
  CK}(\hat{n},t_1+t_2|\hat{m})=Q_{P_s}(\hat{n},t_1+t_2|\hat{m}),\,\forall
t_1,t_2\ge 0$ but
the converse is not true in general
\cite{lapolla2019manifestations,PhysRevResearch.3.L022018}. \textcolor{black}{As
  soon as $Q^{\rm CK}(\hat{n},t_1+t_2|\hat{m})\ne
  Q_{P_s}(\hat{n},t_1+t_2|\hat{m})$ for some $\hat{n}$, the observable at time $t_1+t_2$ ``remembers'' the state of hidden degrees of freedom at time
        $t_1$.}

We use the
Kullback-Leibler divergence
$D_k[p||q]\equiv\sum_kp(k)\ln[p(k)/q(k)]\ge 0$ to
quantify the difference between $Q_{P_s}$  and $Q^{\rm CK}$ \cite{PhysRevResearch.3.L022018} 
\begin{equation}
  D_{\hat{m}}^{\rm CK}(t_1, t_2) \equiv D_{\hat{n}}[Q_{P_s}(\hat{n},t_1+t_2|\hat{m})||Q^{\rm CK}(\hat{n},t_1+t_2|\hat{m})],
\label{KL}  
\end{equation}
where the superscript $k$ in  $D_k[p||q]$ denotes the independent
dummy variable of the measures $p$ and $q$. 
In the absence of memory $D_{\hat{m}}^{\rm CK}(t_1, t_2)=0,\,\forall
t_1,t_2$. Conversely, as we are interested in ergodic dynamics
prepared in a steady state, we have that $D_{\hat{m}}^{\rm CK}\to 0$
whenever $t_1+t_2\to 0$ or $t_1+t_2\to \infty$
\cite{PhysRevResearch.3.L022018}. Therefore, by the positivity of
$D_{\hat{m}}^{\rm CK}$  we will have at least one maximum in the half-space
to $t_1,t_2>0$. We quantify the magnitude of memory in terms of the
\emph{global} maximum on $t_1,t_2>0$ 
\begin{equation}
  D_{\rm max}^{\rm CK}(\hat{m}) \equiv \sup_{t_1,t_2>0}D_{\hat{m}}^{\rm CK}(t_1, t_2).
\label{max}  
\end{equation}
In the baseline setting (\textcolor{black}{$\gamma_1=\gamma_2=1$}) the observed and
hidden dynamics are decoupled (i.e.\ all microscopic
pathways are equivalent). As a result, the
observed dynamics is Markovian and $D_{\rm max}^{\rm CK}(\hat{m})=0$.
We are interested in the dependence of $D_{\rm max}^{\rm CK}(\hat{m})$
as we couple the vertical and horizontal dynamics cooperatively or by
a dissipative current,
that is, on $\gamma_1$ and $\gamma$ in the driven and cooperative
model, respectively.
	\begin{figure}[h]
		\includegraphics[width=0.45\textwidth]{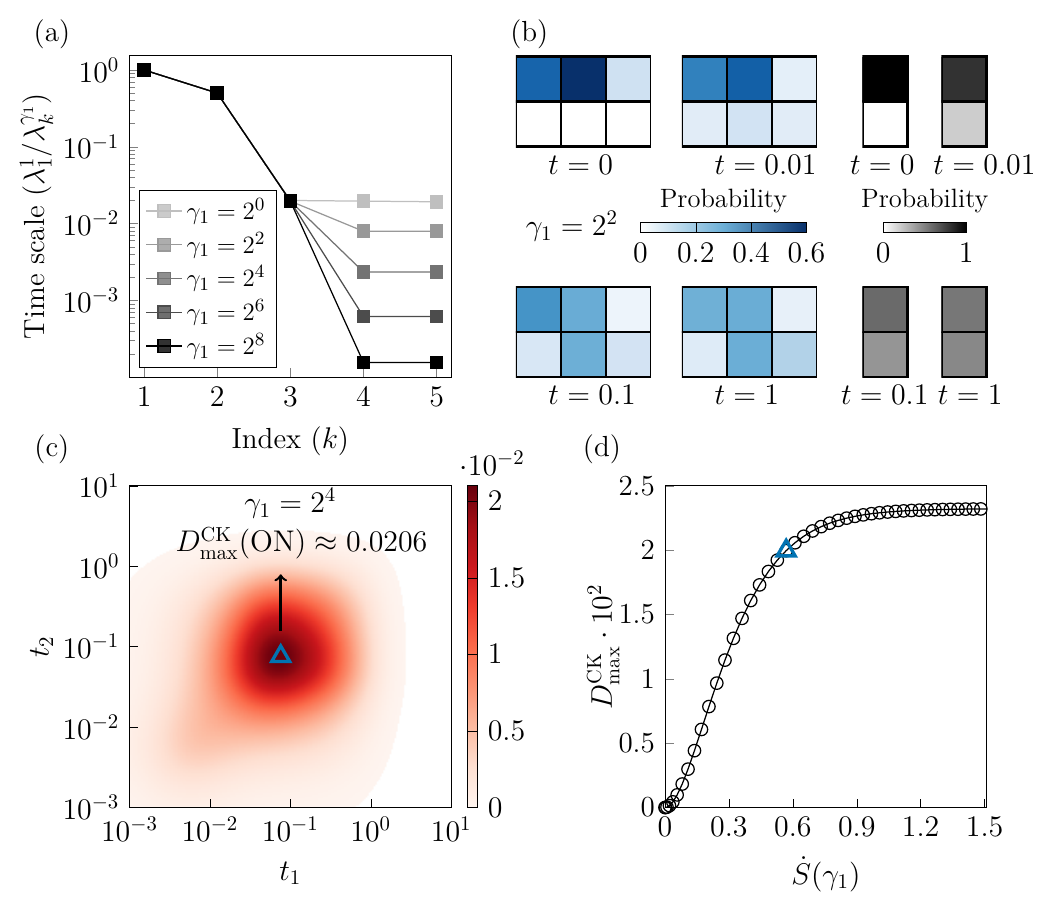}
		\caption{Driven setting ($\gamma_1>1, \gamma_2=1$):\ (a)~Characteristic time-scales
                  $1/\lambda_i^{\gamma_1},\,i\ge 1$ as a function of
                  $\gamma_1$ relative to the
                  baseline relaxation time $1/\lambda^1_1$.~(b) Left:
                  Microscopic transition probability 
                  $G(j,t|{\rm ON})$ for $\gamma_1=2^2$ at four
                  different times. \textcolor{black}{Right: Corresponding observed transition probability $Q_{P_s}(\hat{n},t|\rm ON)$.}~(c)~Relative entropy $D_{\rm ON}^{\rm CK}(t_1, t_2)$
                  in Eq.~\eqref{KL} for $\gamma_1=2^4$;~The triangle depicts $D^{\text{CK}}_{\rm max}({\rm ON})$.~(d)~Magnitude of memory
                  $D^{\text{CK}}_{\rm max}({\rm ON})$ in
                  Eq.~\eqref{max} as a function of driving
                  $\dot{S}(\gamma_1)$;~the blue triangle denotes
                  the position of maximum in (c).}
                \label{fig2}
	\end{figure}
        
\emph{Driven setting.---}We first consider the driven scenario with
$\gamma_1>1$ and $\gamma_2=1$. Instead of $\gamma_1$ we use the
steady-state entropy 
production rate of the microscopic dynamics $\dot{S}(\gamma_1) =
\sum_{i,j} P_s(j) L_{i,j}\ln[P_s(j) L_{i,j}/(P_s(i) L_{j,i})]$ to
indicate how far the system is driven out of equilibrium and we change
$\gamma$ in equidistant units of the chemical potential $\ln (\gamma_1/\gamma_2)=\ln \gamma_1$ that drives the system out of equilibrium, i.e. $\gamma$ increases exponentially.
We first provide some intuition about the microscopic dynamics.

Since $\lambda_{1,2}$ are independent of $\gamma_1$ and because 
$\lambda_3\simeq 2(a+b)$ for $b\gg a$ and $\gamma_1\ge 2$, by increasing
$\gamma_1$ we only alter $\lambda_{4,5}$ (see Fig.~\ref{fig2}a \textcolor{black}{and Appendix B}). That
is, we are essentially only tuning the fastest time scales, whereas
the slow time scales remain unaffected by the driving. We
also alter the stationary distribution $P_s(\hat{m})$. Because $b\gg a$ the system (even without driving) tends to first
explore vertical paths 
to reach a ``quasi-steady state'' between observed ON-OFF states
within a time scale of approximately $\sim 1/b=0.04$ (see
Fig.~\ref{fig2}b). Afterwards,
the probability redistributes horizontally within observable states.
However, because the transition rates from state 1 to 2 and from 6 to
5 are accelerated by a factor of $\gamma_1$, transitions $2\to 1$ and
$6\to 5$ will be instantly followed by the reverse transitions $1\to 2$ and
$5\to 6$. Thus, the probability distribution dominantly redistributes
along the
microscopic path  $2\leftrightarrow4 \leftrightarrow3
\leftrightarrow5$, and finally reaches a steady state
``skewed'' in the hidden direction (see Fig.~\ref{fig2}b). 

We now address the magnitude of the emerging memory via $D^{\rm CK}_{\rm
  max}$ in Eq.~\eqref{max}. We find that  $D^{\rm CK}_{\rm
  max}$ monotonically increases with $\gamma_1$ (note that
$\dot{S}$ is a monotonically increasing function of $\gamma_1$), and eventually
at $\approx 0.02324$, where the location of the supremum approaches
$t_1=t_2\approx 0.08$ as $\gamma_1\to\infty$. Note that the maximal
memory is attained on a time scale that is \emph{longer} than the
local vertical equilibration time $\approx 1/b=0.04$. 
The saturation may be explained by noticing that as
$\gamma_1\to\infty$, only states $2,3,4,5$ have a non-zero probability 
and accelerated paths are almost never traversed. As a result,
$D^{\rm CK}_{\rm  max}$ no longer changes with $\gamma_1$. 
		 
Note that in this driven setting vertical transitions
are always much faster 
than horizontal ones, which
maintains a separation of time scales between
observed  and hidden dynamics. The memory we observe
is thus ``only'' a manifestation of the relaxation of
hidden degrees of freedom. 
\begin{figure}[h]
  \includegraphics[width=0.45\textwidth]{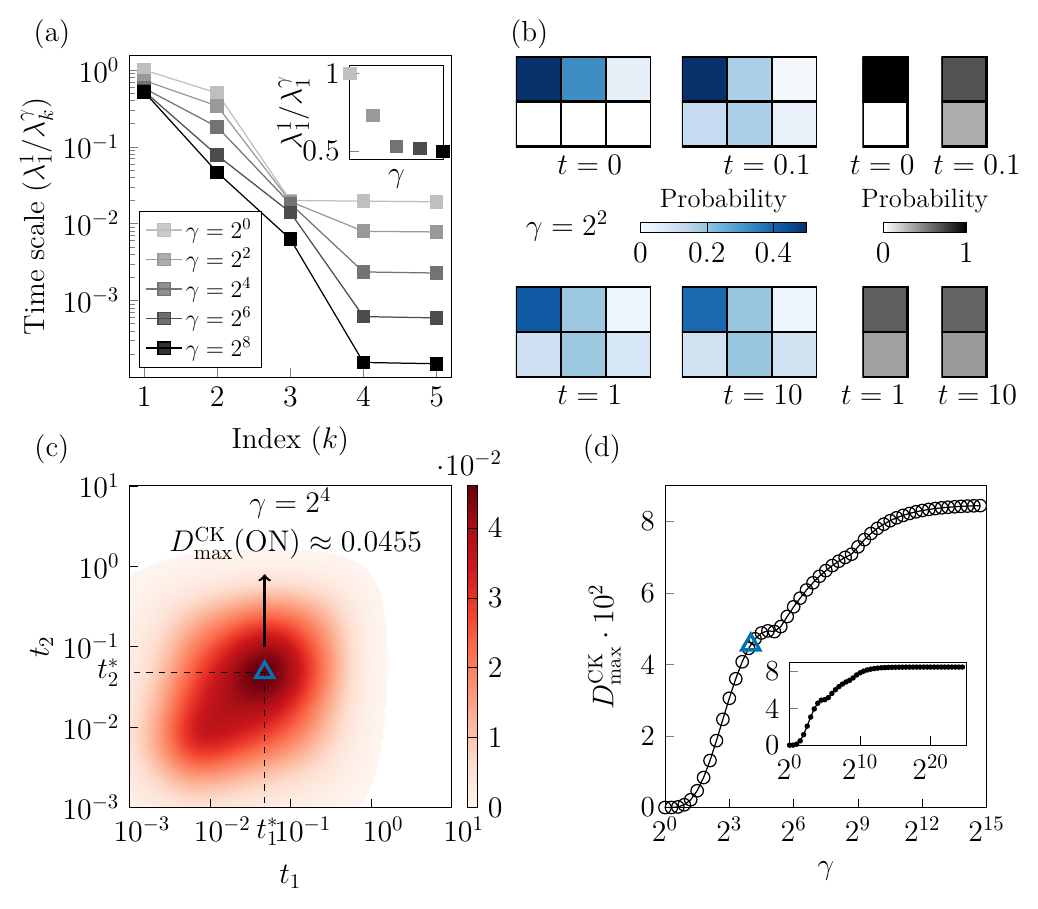}
  \caption{Reversible setting ($\gamma_1=\gamma_2=\gamma$):~(a)~Characteristic time-scales
    $1/\lambda_i^{\gamma},\,i\ge 1$ as a function of
    $\gamma_1=\gamma_2=\gamma$ relative to the
    baseline relaxation time
    $1/\lambda^1_1$.~Inset:~$\lambda^1_1/\lambda_1^{\gamma}$ on a
    linear scale.~(b)~Left: Microscopic transition probability 
                  $G(j,t|{\rm ON})$ for $\gamma=2^2$ at four different
    times. \textcolor{black}{Right: Corresponding observed transition probability $Q_{P_s}(\hat{n},t|\rm ON)$.}~(c)~Relative entropy $D_{\rm ON}^{\rm CK}(t_1, t_2)$
                  in Eq.~\eqref{KL} for $\gamma=2^4$ as a function
                  of $t_1,t_2$;~The triangle denotes
                  $D^{\text{CK}}_{\rm max}({\rm ON})$.~(d)~Magnitude of memory
                  $D^{\text{CK}}_{\rm max}({\rm ON})$ in
                  Eq.~\eqref{max} as a function of $\gamma$.
                  }
  \label{fig3}
\end{figure}
        
\emph{Reversible cooperative setting.---}We now inspect the reversible
scenario where $\gamma_1=\gamma_2\equiv\gamma\ge 1$. As before, we
first give some insight into the microscopic dynamics. As time
evolves, in the first stage the system initially populates microscopic
states 2 and 4, when $\gamma$ is large especially state 2. The
accelerated transition paths do not instantly play a
role. Thus, similar to the driven model, the system tends to first
explore vertical paths (paths $2\to1$ and $4\to 3$) on a time scale of
$1/b\sim 0.04$. In the second stage, also similar to the driven model,
transitions $2\to1$ will instantly go back to state 2. During this stage,
the probability redistributes in the horizontal direction. However,
since the transition rates inside the ON state ($L_{2,4}, L_{4,6}$)
are also accelerated by a factor of $\gamma$, the time scale of the
horizontal redistribution of probability is \emph{not} necessarily
larger than that of vertical dynamics. Here, the two main
frequently visited vertical paths are $2\to 1$ and $4\to 3$.

Note that when we increase $\gamma$ in this reversible
setting, such that $a \gamma > b$, part of the 
horizontal transition rates exceeds vertical ones, 
i.e.\ the time scales of observable dynamics and 
hidden dynamics overlap and there is no time-scale separation.  This
overlapping (and ``mixing'') of hidden and observable time scales may contribute to the
appearance of two shoulders in Fig.~\ref{fig3}d at $\gamma\approx38$
and $\gamma\approx443$. The shoulders are a result of the shift in
position of the peak of $D_{\hat{m}}^{\rm CK}(t_1, t_2)$   (for
details see Fig.~\ref{supp_fig1} in the Appendix A). As $\gamma$ tends to
become very large, $D_{\rm max}^{\rm CK}$ saturates to $\approx
0.08471$, and the location of peak approaches $t_1=t_2\approx 0.13$.

\emph{Conclusion.---}In this Letter we addressed the emergence of memory in 
a minimal setting, 
where the microscopic dynamics corresponds to a Markov process on a planar network and we
observe only the vertical ON$\rightleftarrows$OFF dynamics, whereby 
the  horizontal dynamics are hidden. Our aim was to gain insight into 
how correlations between hidden and observed dynamics emerge, in
particular if and how the nature of these correlations depends on
whether the microscopic dynamics is reversible (i.e.\ obeys detailed
balance) or instead is driven. In the former scenario, the observed and hidden
degrees of freedom are coupled cooperatively, whereas in the latter
scenario the coupling emerges due to a non-equilibrium current. We
focused on quantifying the magnitude of memory while tuning
cooperativity or irreversible driving. Many
features were found to be similar in both setting. However, in the
reversible setting the strongest memory was found in the \emph{expected} situation,
when the time-scales of hidden and observed dynamics overlap. Conversely, in the
driven setting maximal memory is reached under a clear time-scale
separation. Our work therefore unravels qualitative differences in the way memory
can emerge in equilibrium versus driven systems. While we focused on a
simple model, our findings pave the way for more systematic
studies. \textcolor{black}{From a practical, ``\emph{diagnostic}''
  perspective, our results 
imply the possibility to gain insight about the dynamic
coupling underlying active secondary transport~\cite{Tolya_1,Tolya_2,Tolya_3} from 
observations of memory in the transmembrane transport of either species.}

\emph{Acknowledgments.---}Financial support from the Max Planck School
         Matter to Life supported by the German Federal Ministry of
         Education and Research (BMBF) in collaboration with the Max
         Planck Society (fellowship to XZ), and 
         from the European Research Council
(ERC) under the European Union’s Horizon Europe research and
         innovation programme (grant agreement No 101086182 to AG) is
         gratefully acknowledged.
         
\setcounter{equation}{0}
\setcounter{figure}{0}
\setcounter{table}{0}
\renewcommand{\thetable}{B\arabic{table}}
\renewcommand{\thefigure}{A\arabic{figure}}
\renewcommand{\theequation}{A\arabic{equation}}         
\emph{Appendix A:~Peak position of $D^{\rm CK}$ in the reversible setting.---}
Let $(t_1^*,t_2^*)\equiv{\rm arg\,max}_{t_1,t_2}D_{\hat{m}}^{\rm
  CK}(t_1, t_2)$ be the values of $t_1$ and $t_2$ when $D_{\rm ON}^{\rm CK}$
reaches its maximum at a given $\gamma$ in the reversible cooperative
scenario (see Fig.~\ref{fig3}c). As stated in the main text, the
two shoulders in Fig.~\ref{fig3}d are a result of discontinuities in the shift of peak
position. To visualize this, we show in Fig.~\ref{supp_fig1} the
dependence of the peak position on $\gamma$. Note that $G^{\rm CK}$ in
Eq.~\eqref{Gck} is a symmetric function of $t_1$ and $t_2$, so the
peak of $D_{\rm ON}^{\rm CK}$ always occurs at $t_1^* = t_2^*$.
	\begin{figure}[h]
		\includegraphics[width=0.25\textwidth]{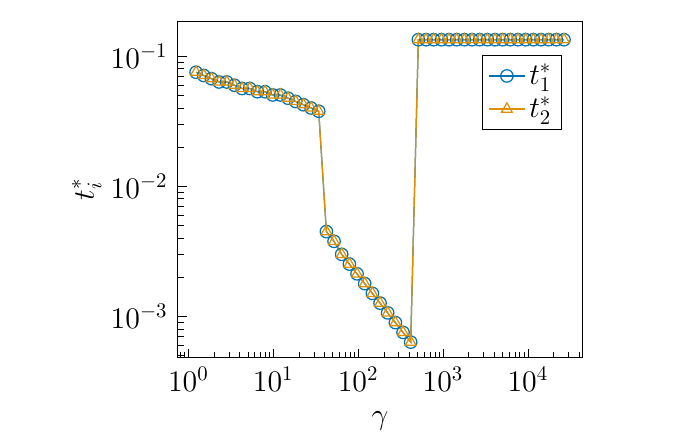}
		\caption{Dependence of the peak position of $D_{\rm ON}^{\rm CK}$,
                  $(t_1^*,t_2^*)$, on the cooperativity parameter
                  $\gamma$ in the reversible
                  model. Note the two discontinuities.}
                \label{supp_fig1}
	\end{figure}

\textcolor{black}{
\emph{Appendix B:~Table of characteristic time-scales.---}
Table \ref{tab:table3} lists a part of the characteristic time-scales
under the driven and reversible settings shown in Figs. \ref{fig2}a
and \ref{fig3}a, respectively.}
\begin{table*}
	\caption{\label{tab:table3}Characteristic time-scales relative
          to the baseline relaxation time in driven ($\lambda_1^1/\lambda_{k}^{\gamma_1}$) and reversible ($\lambda_1^1/\lambda_{k}^{\gamma}$) settings.}
	\begin{ruledtabular}
		\begin{tabular}{cccccc}
			&Baseline&\multicolumn{2}{c}{Driven ($\gamma_1>1, \gamma_2=1$)} &\multicolumn{2}{c}{Reversible ($\gamma_1=\gamma_2=\gamma$)}\\
			Index $k$ &$\gamma_1=\gamma_2=1$&$\gamma_1=2^4$&$\gamma_1=2^8$
			&$\gamma=2^4$&$\gamma=2^8$\\ \hline
			1&1&1 &1&0.578&0.519 \\
			2&0.5&0.5&0.5&0.181&0.0465 \\
			3&0.02&0.0196&0.0196&0.0182&$6.29\times 10^{-3}$\\
			4&0.0196&$2.35\times 10^{-3}$&$1.56\times 10^{-4}$&$2.35\times 10^{-3}$&$1.56\times 10^{-4}$ \\
			5&0.0192 &$2.35\times 10^{-3}$&$1.56\times 10^{-4}$ &$2.27\times 10^{-3}$&$1.50\times 10^{-4}$\\
		\end{tabular}
	\end{ruledtabular}
\end{table*}
	
\bibliography{manuscript_memory}

\begin{thebibliography}{69}%
\makeatletter
\providecommand \@ifxundefined [1]{%
 \@ifx{#1\undefined}
}%
\providecommand \@ifnum [1]{%
 \ifnum #1\expandafter \@firstoftwo
 \else \expandafter \@secondoftwo
 \fi
}%
\providecommand \@ifx [1]{%
 \ifx #1\expandafter \@firstoftwo
 \else \expandafter \@secondoftwo
 \fi
}%
\providecommand \natexlab [1]{#1}%
\providecommand \enquote  [1]{``#1''}%
\providecommand \bibnamefont  [1]{#1}%
\providecommand \bibfnamefont [1]{#1}%
\providecommand \citenamefont [1]{#1}%
\providecommand \href@noop [0]{\@secondoftwo}%
\providecommand \href [0]{\begingroup \@sanitize@url \@href}%
\providecommand \@href[1]{\@@startlink{#1}\@@href}%
\providecommand \@@href[1]{\endgroup#1\@@endlink}%
\providecommand \@sanitize@url [0]{\catcode `\\12\catcode `\$12\catcode
  `\&12\catcode `\#12\catcode `\^12\catcode `\_12\catcode `\%12\relax}%
\providecommand \@@startlink[1]{}%
\providecommand \@@endlink[0]{}%
\providecommand \url  [0]{\begingroup\@sanitize@url \@url }%
\providecommand \@url [1]{\endgroup\@href {#1}{\urlprefix }}%
\providecommand \urlprefix  [0]{URL }%
\providecommand \Eprint [0]{\href }%
\providecommand \doibase [0]{https://doi.org/}%
\providecommand \selectlanguage [0]{\@gobble}%
\providecommand \bibinfo  [0]{\@secondoftwo}%
\providecommand \bibfield  [0]{\@secondoftwo}%
\providecommand \translation [1]{[#1]}%
\providecommand \BibitemOpen [0]{}%
\providecommand \bibitemStop [0]{}%
\providecommand \bibitemNoStop [0]{.\EOS\space}%
\providecommand \EOS [0]{\spacefactor3000\relax}%
\providecommand \BibitemShut  [1]{\csname bibitem#1\endcsname}%
\let\auto@bib@innerbib\@empty
\bibitem [{\citenamefont {Lapolla}\ and\ \citenamefont
  {Godec}(2019)}]{lapolla2019manifestations}%
  \BibitemOpen
  \bibfield  {author} {\bibinfo {author} {\bibfnamefont {A.}~\bibnamefont
  {Lapolla}}\ and\ \bibinfo {author} {\bibfnamefont {A.}~\bibnamefont
  {Godec}},\ }\bibfield  {title} {\bibinfo {title} {Manifestations of
  projection-induced memory: General theory and the tilted single file},\
  }\bibfield  {journal} {\bibinfo  {journal} {Front. Phys.}\ }\textbf {\bibinfo
  {volume} {7}},\ \href {https://doi.org/10.3389/fphy.2019.00182}
  {10.3389/fphy.2019.00182} (\bibinfo {year} {2019})\BibitemShut {NoStop}%
\bibitem [{\citenamefont {Hartich}\ and\ \citenamefont
  {Godec}(2021{\natexlab{a}})}]{PhysRevX.11.041047}%
  \BibitemOpen
  \bibfield  {author} {\bibinfo {author} {\bibfnamefont {D.}~\bibnamefont
  {Hartich}}\ and\ \bibinfo {author} {\bibfnamefont {A.}~\bibnamefont
  {Godec}},\ }\bibfield  {title} {\bibinfo {title} {Emergent memory and kinetic
  hysteresis in strongly driven networks},\ }\href
  {https://doi.org/10.1103/PhysRevX.11.041047} {\bibfield  {journal} {\bibinfo
  {journal} {Phys. Rev. X}\ }\textbf {\bibinfo {volume} {11}},\ \bibinfo
  {pages} {041047} (\bibinfo {year} {2021}{\natexlab{a}})}\BibitemShut
  {NoStop}%
\bibitem [{\citenamefont {Plotkin}\ and\ \citenamefont
  {Wolynes}(1998)}]{Wolynes}%
  \BibitemOpen
  \bibfield  {author} {\bibinfo {author} {\bibfnamefont {S.~S.}\ \bibnamefont
  {Plotkin}}\ and\ \bibinfo {author} {\bibfnamefont {P.~G.}\ \bibnamefont
  {Wolynes}},\ }\bibfield  {title} {\bibinfo {title} {Non-markovian
  configurational diffusion and reaction coordinates for protein folding},\
  }\href {https://doi.org/10.1103/PhysRevLett.80.5015} {\bibfield  {journal}
  {\bibinfo  {journal} {Phys. Rev. Lett.}\ }\textbf {\bibinfo {volume} {80}},\
  \bibinfo {pages} {5015} (\bibinfo {year} {1998})}\BibitemShut {NoStop}%
\bibitem [{\citenamefont {Min}\ \emph {et~al.}(2005)\citenamefont {Min},
  \citenamefont {Luo}, \citenamefont {Cherayil}, \citenamefont {Kou},\ and\
  \citenamefont {Xie}}]{Xie_2005}%
  \BibitemOpen
  \bibfield  {author} {\bibinfo {author} {\bibfnamefont {W.}~\bibnamefont
  {Min}}, \bibinfo {author} {\bibfnamefont {G.}~\bibnamefont {Luo}}, \bibinfo
  {author} {\bibfnamefont {B.~J.}\ \bibnamefont {Cherayil}}, \bibinfo {author}
  {\bibfnamefont {S.~C.}\ \bibnamefont {Kou}},\ and\ \bibinfo {author}
  {\bibfnamefont {X.~S.}\ \bibnamefont {Xie}},\ }\bibfield  {title} {\bibinfo
  {title} {Observation of a power-law memory kernel for fluctuations within a
  single protein molecule},\ }\href
  {https://doi.org/10.1103/PhysRevLett.94.198302} {\bibfield  {journal}
  {\bibinfo  {journal} {Phys. Rev. Lett.}\ }\textbf {\bibinfo {volume} {94}},\
  \bibinfo {pages} {198302} (\bibinfo {year} {2005})}\BibitemShut {NoStop}%
\bibitem [{\citenamefont {Neusius}\ \emph {et~al.}(2008)\citenamefont
  {Neusius}, \citenamefont {Daidone}, \citenamefont {Sokolov},\ and\
  \citenamefont {Smith}}]{JCS_2008}%
  \BibitemOpen
  \bibfield  {author} {\bibinfo {author} {\bibfnamefont {T.}~\bibnamefont
  {Neusius}}, \bibinfo {author} {\bibfnamefont {I.}~\bibnamefont {Daidone}},
  \bibinfo {author} {\bibfnamefont {I.~M.}\ \bibnamefont {Sokolov}},\ and\
  \bibinfo {author} {\bibfnamefont {J.~C.}\ \bibnamefont {Smith}},\ }\bibfield
  {title} {\bibinfo {title} {Subdiffusion in peptides originates from the
  fractal-like structure of configuration space},\ }\href
  {https://doi.org/10.1103/PhysRevLett.100.188103} {\bibfield  {journal}
  {\bibinfo  {journal} {Phys. Rev. Lett.}\ }\textbf {\bibinfo {volume} {100}},\
  \bibinfo {pages} {188103} (\bibinfo {year} {2008})}\BibitemShut {NoStop}%
\bibitem [{\citenamefont {Sangha}\ and\ \citenamefont
  {Keyes}(2009)}]{sangha_proteins_2009}%
  \BibitemOpen
  \bibfield  {author} {\bibinfo {author} {\bibfnamefont {A.~K.}\ \bibnamefont
  {Sangha}}\ and\ \bibinfo {author} {\bibfnamefont {T.}~\bibnamefont {Keyes}},\
  }\bibfield  {title} {\bibinfo {title} {Proteins {Fold} by {Subdiffusion} of
  the {Order} {Parameter}},\ }\href {https://doi.org/10.1021/jp907009r}
  {\bibfield  {journal} {\bibinfo  {journal} {J. Phys. Chem. B}\ }\textbf
  {\bibinfo {volume} {113}},\ \bibinfo {pages} {15886} (\bibinfo {year}
  {2009})}\BibitemShut {NoStop}%
\bibitem [{\citenamefont {Avdoshenko}\ \emph {et~al.}(2017)\citenamefont
  {Avdoshenko}, \citenamefont {Das}, \citenamefont {Satija}, \citenamefont
  {Papoian},\ and\ \citenamefont {Makarov}}]{avdoshenko_theoretical_2017}%
  \BibitemOpen
  \bibfield  {author} {\bibinfo {author} {\bibfnamefont {S.~M.}\ \bibnamefont
  {Avdoshenko}}, \bibinfo {author} {\bibfnamefont {A.}~\bibnamefont {Das}},
  \bibinfo {author} {\bibfnamefont {R.}~\bibnamefont {Satija}}, \bibinfo
  {author} {\bibfnamefont {G.~A.}\ \bibnamefont {Papoian}},\ and\ \bibinfo
  {author} {\bibfnamefont {D.~E.}\ \bibnamefont {Makarov}},\ }\bibfield
  {title} {\bibinfo {title} {Theoretical and computational validation of the
  {Kuhn} barrier friction mechanism in unfolded proteins},\ }\href
  {https://doi.org/10.1038/s41598-017-00287-5} {\bibfield  {journal} {\bibinfo
  {journal} {Sci. Rep.}\ }\textbf {\bibinfo {volume} {7}},\ \bibinfo {pages}
  {269} (\bibinfo {year} {2017})}\BibitemShut {NoStop}%
\bibitem [{\citenamefont {Makarov}(2013)}]{Dima_2013}%
  \BibitemOpen
  \bibfield  {author} {\bibinfo {author} {\bibfnamefont {D.~E.}\ \bibnamefont
  {Makarov}},\ }\bibfield  {title} {\bibinfo {title} {Interplay of non-markov
  and internal friction effects in the barrier crossing kinetics of
  biopolymers: Insights from an analytically solvable model},\ }\href
  {https://doi.org/10.1063/1.4773283} {\bibfield  {journal} {\bibinfo
  {journal} {J. Chem. Phys.}\ }\textbf {\bibinfo {volume} {138}},\ \bibinfo
  {pages} {014102} (\bibinfo {year} {2013})}\BibitemShut {NoStop}%
\bibitem [{\citenamefont {Ozmaian}\ and\ \citenamefont
  {Makarov}(2019)}]{Dima_2019}%
  \BibitemOpen
  \bibfield  {author} {\bibinfo {author} {\bibfnamefont {M.}~\bibnamefont
  {Ozmaian}}\ and\ \bibinfo {author} {\bibfnamefont {D.~E.}\ \bibnamefont
  {Makarov}},\ }\bibfield  {title} {\bibinfo {title} {Transition path dynamics
  in the binding of intrinsically disordered proteins: A simulation study},\
  }\href {https://doi.org/10.1063/1.5129150} {\bibfield  {journal} {\bibinfo
  {journal} {J. Chem. Phys.}\ }\textbf {\bibinfo {volume} {151}},\ \bibinfo
  {pages} {235101} (\bibinfo {year} {2019})}\BibitemShut {NoStop}%
\bibitem [{\citenamefont {Meyer}\ \emph {et~al.}(2020)\citenamefont {Meyer},
  \citenamefont {Pelagejcev},\ and\ \citenamefont {Schilling}}]{Meyer_2020}%
  \BibitemOpen
  \bibfield  {author} {\bibinfo {author} {\bibfnamefont {H.}~\bibnamefont
  {Meyer}}, \bibinfo {author} {\bibfnamefont {P.}~\bibnamefont {Pelagejcev}},\
  and\ \bibinfo {author} {\bibfnamefont {T.}~\bibnamefont {Schilling}},\
  }\bibfield  {title} {\bibinfo {title} {Non-markovian out-of-equilibrium
  dynamics: A general numerical procedure to construct time-dependent memory
  kernels for coarse-grained observables},\ }\href
  {https://doi.org/10.1209/0295-5075/128/40001} {\bibfield  {journal} {\bibinfo
   {journal} {EPL (Europhys. Lett.)}\ }\textbf {\bibinfo {volume} {128}},\
  \bibinfo {pages} {40001} (\bibinfo {year} {2020})}\BibitemShut {NoStop}%
\bibitem [{\citenamefont {Pyo}\ and\ \citenamefont
  {Woodside}(2019)}]{woodside}%
  \BibitemOpen
  \bibfield  {author} {\bibinfo {author} {\bibfnamefont {A.~G.~T.}\
  \bibnamefont {Pyo}}\ and\ \bibinfo {author} {\bibfnamefont {M.~T.}\
  \bibnamefont {Woodside}},\ }\bibfield  {title} {\bibinfo {title} {Memory
  effects in single-molecule force spectroscopy measurements of biomolecular
  folding},\ }\href {https://doi.org/10.1039/c9cp04197d} {\bibfield  {journal}
  {\bibinfo  {journal} {Phys. Chem. Chem. Phys.}\ }\textbf {\bibinfo {volume}
  {21}},\ \bibinfo {pages} {24527–24534} (\bibinfo {year}
  {2019})}\BibitemShut {NoStop}%
\bibitem [{\citenamefont {Ayaz}\ \emph {et~al.}(2022)\citenamefont {Ayaz},
  \citenamefont {Scalfi}, \citenamefont {Dalton},\ and\ \citenamefont
  {Netz}}]{Netz}%
  \BibitemOpen
  \bibfield  {author} {\bibinfo {author} {\bibfnamefont {C.}~\bibnamefont
  {Ayaz}}, \bibinfo {author} {\bibfnamefont {L.}~\bibnamefont {Scalfi}},
  \bibinfo {author} {\bibfnamefont {B.~A.}\ \bibnamefont {Dalton}},\ and\
  \bibinfo {author} {\bibfnamefont {R.~R.}\ \bibnamefont {Netz}},\ }\bibfield
  {title} {\bibinfo {title} {Generalized langevin equation with a nonlinear
  potential of mean force and nonlinear memory friction from a hybrid
  projection scheme},\ }\href {https://doi.org/10.1103/PhysRevE.105.054138}
  {\bibfield  {journal} {\bibinfo  {journal} {Phys. Rev. E}\ }\textbf {\bibinfo
  {volume} {105}},\ \bibinfo {pages} {054138} (\bibinfo {year}
  {2022})}\BibitemShut {NoStop}%
\bibitem [{\citenamefont {Ginot}\ \emph {et~al.}(2022)\citenamefont {Ginot},
  \citenamefont {Caspers}, \citenamefont {Kr\"uger},\ and\ \citenamefont
  {Bechinger}}]{Krueger}%
  \BibitemOpen
  \bibfield  {author} {\bibinfo {author} {\bibfnamefont {F.}~\bibnamefont
  {Ginot}}, \bibinfo {author} {\bibfnamefont {J.}~\bibnamefont {Caspers}},
  \bibinfo {author} {\bibfnamefont {M.}~\bibnamefont {Kr\"uger}},\ and\
  \bibinfo {author} {\bibfnamefont {C.}~\bibnamefont {Bechinger}},\ }\bibfield
  {title} {\bibinfo {title} {Barrier crossing in a viscoelastic bath},\ }\href
  {https://doi.org/10.1103/PhysRevLett.128.028001} {\bibfield  {journal}
  {\bibinfo  {journal} {Phys. Rev. Lett.}\ }\textbf {\bibinfo {volume} {128}},\
  \bibinfo {pages} {028001} (\bibinfo {year} {2022})}\BibitemShut {NoStop}%
\bibitem [{\citenamefont {Narinder}\ \emph {et~al.}(2018)\citenamefont
  {Narinder}, \citenamefont {Bechinger},\ and\ \citenamefont
  {Gomez-Solano}}]{Solano}%
  \BibitemOpen
  \bibfield  {author} {\bibinfo {author} {\bibfnamefont {N.}~\bibnamefont
  {Narinder}}, \bibinfo {author} {\bibfnamefont {C.}~\bibnamefont
  {Bechinger}},\ and\ \bibinfo {author} {\bibfnamefont {J.~R.}\ \bibnamefont
  {Gomez-Solano}},\ }\bibfield  {title} {\bibinfo {title} {Memory-induced
  transition from a persistent random walk to circular motion for achiral
  microswimmers},\ }\href {https://doi.org/10.1103/PhysRevLett.121.078003}
  {\bibfield  {journal} {\bibinfo  {journal} {Phys. Rev. Lett.}\ }\textbf
  {\bibinfo {volume} {121}},\ \bibinfo {pages} {078003} (\bibinfo {year}
  {2018})}\BibitemShut {NoStop}%
\bibitem [{\citenamefont {Haken}(1975)}]{Haken}%
  \BibitemOpen
  \bibfield  {author} {\bibinfo {author} {\bibfnamefont {H.}~\bibnamefont
  {Haken}},\ }\bibfield  {title} {\bibinfo {title} {Cooperative phenomena in
  systems far from thermal equilibrium and in nonphysical systems},\ }\href
  {https://doi.org/10.1103/RevModPhys.47.67} {\bibfield  {journal} {\bibinfo
  {journal} {Rev. Mod. Phys.}\ }\textbf {\bibinfo {volume} {47}},\ \bibinfo
  {pages} {67} (\bibinfo {year} {1975})}\BibitemShut {NoStop}%
\bibitem [{\citenamefont {Grabert}\ \emph {et~al.}(1977)\citenamefont
  {Grabert}, \citenamefont {Talkner},\ and\ \citenamefont
  {H{\"a}nggi}}]{Haenggi_1}%
  \BibitemOpen
  \bibfield  {author} {\bibinfo {author} {\bibfnamefont {H.}~\bibnamefont
  {Grabert}}, \bibinfo {author} {\bibfnamefont {P.}~\bibnamefont {Talkner}},\
  and\ \bibinfo {author} {\bibfnamefont {P.}~\bibnamefont {H{\"a}nggi}},\
  }\bibfield  {title} {\bibinfo {title} {Microdynamics and time-evolution of
  macroscopic non-markovian systems},\ }\href
  {http://dx.doi.org/10.1007/BF01570749} {\bibfield  {journal} {\bibinfo
  {journal} {Z. Phys. B: Cond. Matt.}\ }\textbf {\bibinfo {volume} {26}},\
  \bibinfo {pages} {389} (\bibinfo {year} {1977})}\BibitemShut {NoStop}%
\bibitem [{\citenamefont {Grabert}\ \emph {et~al.}(1978)\citenamefont
  {Grabert}, \citenamefont {Talkner}, \citenamefont {Hänggi},\ and\
  \citenamefont {Thomas}}]{Haenggi_2}%
  \BibitemOpen
  \bibfield  {author} {\bibinfo {author} {\bibfnamefont {H.}~\bibnamefont
  {Grabert}}, \bibinfo {author} {\bibfnamefont {P.}~\bibnamefont {Talkner}},
  \bibinfo {author} {\bibfnamefont {P.}~\bibnamefont {Hänggi}},\ and\ \bibinfo
  {author} {\bibfnamefont {H.}~\bibnamefont {Thomas}},\ }\bibfield  {title}
  {\bibinfo {title} {Microdynamics and time-evolution of macroscopic
  non-markovian systems. ii},\ }\href {https://doi.org/10.1007/bf01321192}
  {\bibfield  {journal} {\bibinfo  {journal} {Z. Phys. B: Cond. Matt.}\
  }\textbf {\bibinfo {volume} {29}},\ \bibinfo {pages} {273–280} (\bibinfo
  {year} {1978})}\BibitemShut {NoStop}%
\bibitem [{\citenamefont {Casademunt}\ \emph {et~al.}(1987)\citenamefont
  {Casademunt}, \citenamefont {Mannella}, \citenamefont {McClintock},
  \citenamefont {Moss},\ and\ \citenamefont {Sancho}}]{Sancho}%
  \BibitemOpen
  \bibfield  {author} {\bibinfo {author} {\bibfnamefont {J.}~\bibnamefont
  {Casademunt}}, \bibinfo {author} {\bibfnamefont {R.}~\bibnamefont
  {Mannella}}, \bibinfo {author} {\bibfnamefont {P.~V.~E.}\ \bibnamefont
  {McClintock}}, \bibinfo {author} {\bibfnamefont {F.~E.}\ \bibnamefont
  {Moss}},\ and\ \bibinfo {author} {\bibfnamefont {J.~M.}\ \bibnamefont
  {Sancho}},\ }\bibfield  {title} {\bibinfo {title} {Relaxation times of
  non-markovian processes},\ }\href {https://doi.org/10.1103/PhysRevA.35.5183}
  {\bibfield  {journal} {\bibinfo  {journal} {Phys. Rev. A}\ }\textbf {\bibinfo
  {volume} {35}},\ \bibinfo {pages} {5183} (\bibinfo {year}
  {1987})}\BibitemShut {NoStop}%
\bibitem [{\citenamefont {Ferrario}\ and\ \citenamefont
  {Grigolini}(2008)}]{Grigolini}%
  \BibitemOpen
  \bibfield  {author} {\bibinfo {author} {\bibfnamefont {M.}~\bibnamefont
  {Ferrario}}\ and\ \bibinfo {author} {\bibfnamefont {P.}~\bibnamefont
  {Grigolini}},\ }\bibfield  {title} {\bibinfo {title} {{The non‐Markovian
  relaxation process as a ‘‘contraction’’ of a multidimensional one of
  Markovian type}},\ }\href {https://doi.org/10.1063/1.524019} {\bibfield
  {journal} {\bibinfo  {journal} {J. Math. Phys.}\ }\textbf {\bibinfo {volume}
  {20}},\ \bibinfo {pages} {2567} (\bibinfo {year} {2008})}\BibitemShut
  {NoStop}%
\bibitem [{\citenamefont {Goychuk}(2012)}]{Goychuk}%
  \BibitemOpen
  \bibfield  {author} {\bibinfo {author} {\bibfnamefont {I.}~\bibnamefont
  {Goychuk}},\ }\bibfield  {title} {\bibinfo {title} {Viscoelastic
  subdiffusion: Generalized langevin equation approach},\ }\href
  {https://doi.org/10.1002/9781118197714.ch5} {\bibfield  {journal} {\bibinfo
  {journal} {Adv. Chem. Phys.}\ ,\ \bibinfo {pages} {187–253}} (\bibinfo
  {year} {2012})}\BibitemShut {NoStop}%
\bibitem [{\citenamefont {Sokolov}(2002)}]{Igor}%
  \BibitemOpen
  \bibfield  {author} {\bibinfo {author} {\bibfnamefont {I.~M.}\ \bibnamefont
  {Sokolov}},\ }\bibfield  {title} {\bibinfo {title} {Solutions of a class of
  non-{M}arkovian {F}okker-{P}lanck equations},\ }\href
  {https://doi.org/10.1103/PhysRevE.66.041101} {\bibfield  {journal} {\bibinfo
  {journal} {Phys. Rev. E}\ }\textbf {\bibinfo {volume} {66}},\ \bibinfo
  {pages} {041101} (\bibinfo {year} {2002})}\BibitemShut {NoStop}%
\bibitem [{\citenamefont {Sokolov}(2003)}]{Igor_2}%
  \BibitemOpen
  \bibfield  {author} {\bibinfo {author} {\bibfnamefont {I.~M.}\ \bibnamefont
  {Sokolov}},\ }\bibfield  {title} {\bibinfo {title} {Cyclization of a polymer:
  First-passage problem for a {N}on-{M}arkovian process},\ }\href
  {https://doi.org/10.1103/PhysRevLett.90.080601} {\bibfield  {journal}
  {\bibinfo  {journal} {Phys. Rev. Lett.}\ }\textbf {\bibinfo {volume} {90}},\
  \bibinfo {pages} {080601} (\bibinfo {year} {2003})}\BibitemShut {NoStop}%
\bibitem [{\citenamefont {Dybiec}\ \emph {et~al.}(2011)\citenamefont {Dybiec},
  \citenamefont {Sokolov},\ and\ \citenamefont {Chechkin}}]{Igor_3}%
  \BibitemOpen
  \bibfield  {author} {\bibinfo {author} {\bibfnamefont {B.}~\bibnamefont
  {Dybiec}}, \bibinfo {author} {\bibfnamefont {I.~M.}\ \bibnamefont
  {Sokolov}},\ and\ \bibinfo {author} {\bibfnamefont {A.~V.}\ \bibnamefont
  {Chechkin}},\ }\bibfield  {title} {\bibinfo {title} {Relaxation to stationary
  states for anomalous diffusion},\ }\href
  {https://doi.org/10.1016/j.cnsns.2011.05.011} {\bibfield  {journal} {\bibinfo
   {journal} {Commun. Nonlinear Sci. Numer. Simul.}\ }\textbf {\bibinfo
  {volume} {16}},\ \bibinfo {pages} {4549–4557} (\bibinfo {year}
  {2011})}\BibitemShut {NoStop}%
\bibitem [{\citenamefont {Metzler}\ and\ \citenamefont
  {Klafter}(2000)}]{Metz_1}%
  \BibitemOpen
  \bibfield  {author} {\bibinfo {author} {\bibfnamefont {R.}~\bibnamefont
  {Metzler}}\ and\ \bibinfo {author} {\bibfnamefont {J.}~\bibnamefont
  {Klafter}},\ }\bibfield  {title} {\bibinfo {title} {The random walk's guide
  to anomalous diffusion: a fractional dynamics approach},\ }\href
  {https://doi.org/https://doi.org/10.1016/S0370-1573(00)00070-3} {\bibfield
  {journal} {\bibinfo  {journal} {Phys. Rep.}\ }\textbf {\bibinfo {volume}
  {339}},\ \bibinfo {pages} {1} (\bibinfo {year} {2000})}\BibitemShut {NoStop}%
\bibitem [{\citenamefont {Metzler}\ and\ \citenamefont
  {Klafter}(2004)}]{Metz_2}%
  \BibitemOpen
  \bibfield  {author} {\bibinfo {author} {\bibfnamefont {R.}~\bibnamefont
  {Metzler}}\ and\ \bibinfo {author} {\bibfnamefont {J.}~\bibnamefont
  {Klafter}},\ }\bibfield  {title} {\bibinfo {title} {The restaurant at the end
  of the random walk: recent developments in the description of anomalous
  transport by fractional dynamics},\ }\href
  {https://doi.org/10.1088/0305-4470/37/31/R01} {\bibfield  {journal} {\bibinfo
   {journal} {J. Phys. A: Math. Gen.}\ }\textbf {\bibinfo {volume} {37}},\
  \bibinfo {pages} {R161} (\bibinfo {year} {2004})}\BibitemShut {NoStop}%
\bibitem [{\citenamefont {Barkai}\ \emph {et~al.}(2000)\citenamefont {Barkai},
  \citenamefont {Metzler},\ and\ \citenamefont {Klafter}}]{Eli_1}%
  \BibitemOpen
  \bibfield  {author} {\bibinfo {author} {\bibfnamefont {E.}~\bibnamefont
  {Barkai}}, \bibinfo {author} {\bibfnamefont {R.}~\bibnamefont {Metzler}},\
  and\ \bibinfo {author} {\bibfnamefont {J.}~\bibnamefont {Klafter}},\
  }\bibfield  {title} {\bibinfo {title} {From continuous time random walks to
  the fractional fokker-planck equation},\ }\href
  {https://doi.org/10.1103/PhysRevE.61.132} {\bibfield  {journal} {\bibinfo
  {journal} {Phys. Rev. E}\ }\textbf {\bibinfo {volume} {61}},\ \bibinfo
  {pages} {132} (\bibinfo {year} {2000})}\BibitemShut {NoStop}%
\bibitem [{\citenamefont {Schulz}\ \emph {et~al.}(2014)\citenamefont {Schulz},
  \citenamefont {Barkai},\ and\ \citenamefont {Metzler}}]{Eli_2}%
  \BibitemOpen
  \bibfield  {author} {\bibinfo {author} {\bibfnamefont {J.~H.~P.}\
  \bibnamefont {Schulz}}, \bibinfo {author} {\bibfnamefont {E.}~\bibnamefont
  {Barkai}},\ and\ \bibinfo {author} {\bibfnamefont {R.}~\bibnamefont
  {Metzler}},\ }\bibfield  {title} {\bibinfo {title} {Aging renewal theory and
  application to random walks},\ }\href
  {https://doi.org/10.1103/PhysRevX.4.011028} {\bibfield  {journal} {\bibinfo
  {journal} {Phys. Rev. X}\ }\textbf {\bibinfo {volume} {4}},\ \bibinfo {pages}
  {011028} (\bibinfo {year} {2014})}\BibitemShut {NoStop}%
\bibitem [{\citenamefont {Carmi}\ \emph {et~al.}(2010)\citenamefont {Carmi},
  \citenamefont {Turgeman},\ and\ \citenamefont {Barkai}}]{Eli_3}%
  \BibitemOpen
  \bibfield  {author} {\bibinfo {author} {\bibfnamefont {S.}~\bibnamefont
  {Carmi}}, \bibinfo {author} {\bibfnamefont {L.}~\bibnamefont {Turgeman}},\
  and\ \bibinfo {author} {\bibfnamefont {E.}~\bibnamefont {Barkai}},\
  }\bibfield  {title} {\bibinfo {title} {On distributions of functionals of
  anomalous diffusion paths},\ }\href
  {https://doi.org/10.1007/s10955-010-0086-6} {\bibfield  {journal} {\bibinfo
  {journal} {J. Stat. Phys.}\ }\textbf {\bibinfo {volume} {141}},\ \bibinfo
  {pages} {1071–1092} (\bibinfo {year} {2010})}\BibitemShut {NoStop}%
\bibitem [{\citenamefont {Barkai}\ and\ \citenamefont
  {Sokolov}(2007)}]{Eli_Igor}%
  \BibitemOpen
  \bibfield  {author} {\bibinfo {author} {\bibfnamefont {E.}~\bibnamefont
  {Barkai}}\ and\ \bibinfo {author} {\bibfnamefont {I.~M.}\ \bibnamefont
  {Sokolov}},\ }\bibfield  {title} {\bibinfo {title} {Multi-point distribution
  function for the continuous time random walk},\ }\href
  {https://doi.org/10.1088/1742-5468/2007/08/P08001} {\bibfield  {journal}
  {\bibinfo  {journal} {J. Stat. Mech.}\ }\textbf {\bibinfo {volume} {2007}},\
  \bibinfo {pages} {P08001} (\bibinfo {year} {2007})}\BibitemShut {NoStop}%
\bibitem [{\citenamefont {Bray}\ \emph {et~al.}(2013)\citenamefont {Bray},
  \citenamefont {Majumdar},\ and\ \citenamefont {Schehr}}]{Satya2}%
  \BibitemOpen
  \bibfield  {author} {\bibinfo {author} {\bibfnamefont {A.~J.}\ \bibnamefont
  {Bray}}, \bibinfo {author} {\bibfnamefont {S.~N.}\ \bibnamefont {Majumdar}},\
  and\ \bibinfo {author} {\bibfnamefont {G.}~\bibnamefont {Schehr}},\
  }\bibfield  {title} {\bibinfo {title} {Persistence and first-passage
  properties in nonequilibrium systems},\ }\href
  {https://doi.org/10.1080/00018732.2013.803819} {\bibfield  {journal}
  {\bibinfo  {journal} {Adv. Phys.}\ }\textbf {\bibinfo {volume} {62}},\
  \bibinfo {pages} {225} (\bibinfo {year} {2013})},\ \Eprint
  {https://arxiv.org/abs/https://doi.org/10.1080/00018732.2013.803819}
  {https://doi.org/10.1080/00018732.2013.803819} \BibitemShut {NoStop}%
\bibitem [{\citenamefont {Majumdar}\ and\ \citenamefont
  {Bray}(1998)}]{Satya_c1}%
  \BibitemOpen
  \bibfield  {author} {\bibinfo {author} {\bibfnamefont {S.~N.}\ \bibnamefont
  {Majumdar}}\ and\ \bibinfo {author} {\bibfnamefont {A.~J.}\ \bibnamefont
  {Bray}},\ }\bibfield  {title} {\bibinfo {title} {Persistence with partial
  survival},\ }\href {https://doi.org/10.1103/PhysRevLett.81.2626} {\bibfield
  {journal} {\bibinfo  {journal} {Phys. Rev. Lett.}\ }\textbf {\bibinfo
  {volume} {81}},\ \bibinfo {pages} {2626} (\bibinfo {year}
  {1998})}\BibitemShut {NoStop}%
\bibitem [{\citenamefont {Majumdar}\ \emph {et~al.}(1996)\citenamefont
  {Majumdar}, \citenamefont {Bray}, \citenamefont {Cornell},\ and\
  \citenamefont {Sire}}]{Satya_critical}%
  \BibitemOpen
  \bibfield  {author} {\bibinfo {author} {\bibfnamefont {S.~N.}\ \bibnamefont
  {Majumdar}}, \bibinfo {author} {\bibfnamefont {A.~J.}\ \bibnamefont {Bray}},
  \bibinfo {author} {\bibfnamefont {S.~J.}\ \bibnamefont {Cornell}},\ and\
  \bibinfo {author} {\bibfnamefont {C.}~\bibnamefont {Sire}},\ }\bibfield
  {title} {\bibinfo {title} {Global persistence exponent for nonequilibrium
  critical dynamics},\ }\href {https://doi.org/10.1103/PhysRevLett.77.3704}
  {\bibfield  {journal} {\bibinfo  {journal} {Phys. Rev. Lett.}\ }\textbf
  {\bibinfo {volume} {77}},\ \bibinfo {pages} {3704} (\bibinfo {year}
  {1996})}\BibitemShut {NoStop}%
\bibitem [{\citenamefont {Lapolla}\ and\ \citenamefont
  {Godec}(2020)}]{Lapolla_2020}%
  \BibitemOpen
  \bibfield  {author} {\bibinfo {author} {\bibfnamefont {A.}~\bibnamefont
  {Lapolla}}\ and\ \bibinfo {author} {\bibfnamefont {A.}~\bibnamefont
  {Godec}},\ }\bibfield  {title} {\bibinfo {title} {Single-file diffusion in a
  bi-stable potential: Signatures of memory in the barrier-crossing of a
  tagged-particle},\ }\href {https://doi.org/10.1063/5.0025785} {\bibfield
  {journal} {\bibinfo  {journal} {J. Chem. Phys.}\ }\textbf {\bibinfo {volume}
  {153}},\ \bibinfo {pages} {194104} (\bibinfo {year} {2020})}\BibitemShut
  {NoStop}%
\bibitem [{\citenamefont {Talkner}\ and\ \citenamefont
  {H\"anggi}(2020)}]{Haenggi_RMP}%
  \BibitemOpen
  \bibfield  {author} {\bibinfo {author} {\bibfnamefont {P.}~\bibnamefont
  {Talkner}}\ and\ \bibinfo {author} {\bibfnamefont {P.}~\bibnamefont
  {H\"anggi}},\ }\bibfield  {title} {\bibinfo {title} {Colloquium: Statistical
  mechanics and thermodynamics at strong coupling: Quantum and classical},\
  }\href {https://doi.org/10.1103/RevModPhys.92.041002} {\bibfield  {journal}
  {\bibinfo  {journal} {Rev. Mod. Phys.}\ }\textbf {\bibinfo {volume} {92}},\
  \bibinfo {pages} {041002} (\bibinfo {year} {2020})}\BibitemShut {NoStop}%
\bibitem [{\citenamefont {Di~Terlizzi}\ \emph {et~al.}(2020)\citenamefont
  {Di~Terlizzi}, \citenamefont {Ritort},\ and\ \citenamefont
  {Baiesi}}]{Baiesi_GLE}%
  \BibitemOpen
  \bibfield  {author} {\bibinfo {author} {\bibfnamefont {I.}~\bibnamefont
  {Di~Terlizzi}}, \bibinfo {author} {\bibfnamefont {F.}~\bibnamefont
  {Ritort}},\ and\ \bibinfo {author} {\bibfnamefont {M.}~\bibnamefont
  {Baiesi}},\ }\bibfield  {title} {\bibinfo {title} {Explicit solution of the
  generalised langevin equation},\ }\href
  {https://doi.org/10.1007/s10955-020-02639-4} {\bibfield  {journal} {\bibinfo
  {journal} {J. Stat. Phys.}\ }\textbf {\bibinfo {volume} {181}},\ \bibinfo
  {pages} {1609–1635} (\bibinfo {year} {2020})}\BibitemShut {NoStop}%
\bibitem [{\citenamefont {Netz}(2023{\natexlab{a}})}]{netz2023derivation}%
  \BibitemOpen
  \bibfield  {author} {\bibinfo {author} {\bibfnamefont {R.~R.}\ \bibnamefont
  {Netz}},\ }\href@noop {} {\bibinfo {title} {Derivation of the non-equilibrium
  generalized langevin equation from a generic time-dependent hamiltonian}}
  (\bibinfo {year} {2023}{\natexlab{a}}),\ \Eprint
  {https://arxiv.org/abs/2310.00748} {arXiv:2310.00748 [cond-mat.stat-mech]}
  \BibitemShut {NoStop}%
\bibitem [{\citenamefont {Netz}(2023{\natexlab{b}})}]{netz2023multipoint}%
  \BibitemOpen
  \bibfield  {author} {\bibinfo {author} {\bibfnamefont {R.~R.}\ \bibnamefont
  {Netz}},\ }\href@noop {} {\bibinfo {title} {Multi-point distribution for
  gaussian non-equilibrium non-markovian observables}} (\bibinfo {year}
  {2023}{\natexlab{b}}),\ \Eprint {https://arxiv.org/abs/2310.08886}
  {arXiv:2310.08886 [cond-mat.stat-mech]} \BibitemShut {NoStop}%
\bibitem [{\citenamefont {Terlizzi}\ and\ \citenamefont
  {Baiesi}(2020)}]{Baiesi_TUR}%
  \BibitemOpen
  \bibfield  {author} {\bibinfo {author} {\bibfnamefont {I.~D.}\ \bibnamefont
  {Terlizzi}}\ and\ \bibinfo {author} {\bibfnamefont {M.}~\bibnamefont
  {Baiesi}},\ }\bibfield  {title} {\bibinfo {title} {A thermodynamic
  uncertainty relation for a system with memory},\ }\href
  {https://doi.org/10.1088/1751-8121/abbc7d} {\bibfield  {journal} {\bibinfo
  {journal} {J. Phys. A: Math. Theor.}\ }\textbf {\bibinfo {volume} {53}},\
  \bibinfo {pages} {474002} (\bibinfo {year} {2020})}\BibitemShut {NoStop}%
\bibitem [{\citenamefont {Mehl}\ \emph {et~al.}(2012)\citenamefont {Mehl},
  \citenamefont {Lander}, \citenamefont {Bechinger}, \citenamefont {Blickle},\
  and\ \citenamefont {Seifert}}]{PhysRevLett.108.220601}%
  \BibitemOpen
  \bibfield  {author} {\bibinfo {author} {\bibfnamefont {J.}~\bibnamefont
  {Mehl}}, \bibinfo {author} {\bibfnamefont {B.}~\bibnamefont {Lander}},
  \bibinfo {author} {\bibfnamefont {C.}~\bibnamefont {Bechinger}}, \bibinfo
  {author} {\bibfnamefont {V.}~\bibnamefont {Blickle}},\ and\ \bibinfo {author}
  {\bibfnamefont {U.}~\bibnamefont {Seifert}},\ }\bibfield  {title} {\bibinfo
  {title} {Role of hidden slow degrees of freedom in the fluctuation theorem},\
  }\href {https://doi.org/10.1103/PhysRevLett.108.220601} {\bibfield  {journal}
  {\bibinfo  {journal} {Phys. Rev. Lett.}\ }\textbf {\bibinfo {volume} {108}},\
  \bibinfo {pages} {220601} (\bibinfo {year} {2012})}\BibitemShut {NoStop}%
\bibitem [{\citenamefont {Esposito}(2012)}]{Massi}%
  \BibitemOpen
  \bibfield  {author} {\bibinfo {author} {\bibfnamefont {M.}~\bibnamefont
  {Esposito}},\ }\bibfield  {title} {\bibinfo {title} {Stochastic
  thermodynamics under coarse graining},\ }\href
  {https://doi.org/10.1103/PhysRevE.85.041125} {\bibfield  {journal} {\bibinfo
  {journal} {Phys. Rev. E}\ }\textbf {\bibinfo {volume} {85}},\ \bibinfo
  {pages} {041125} (\bibinfo {year} {2012})}\BibitemShut {NoStop}%
\bibitem [{\citenamefont {van~der Meer}\ \emph {et~al.}(2022)\citenamefont
  {van~der Meer}, \citenamefont {Ertel},\ and\ \citenamefont
  {Seifert}}]{van2022thermodynamic}%
  \BibitemOpen
  \bibfield  {author} {\bibinfo {author} {\bibfnamefont {J.}~\bibnamefont
  {van~der Meer}}, \bibinfo {author} {\bibfnamefont {B.}~\bibnamefont
  {Ertel}},\ and\ \bibinfo {author} {\bibfnamefont {U.}~\bibnamefont
  {Seifert}},\ }\bibfield  {title} {\bibinfo {title} {Thermodynamic inference
  in partially accessible markov networks: A unifying perspective from
  transition-based waiting time distributions},\ }\href
  {https://doi.org/10.1103/PhysRevX.12.031025} {\bibfield  {journal} {\bibinfo
  {journal} {Phys. Rev. X}\ }\textbf {\bibinfo {volume} {12}},\ \bibinfo
  {pages} {031025} (\bibinfo {year} {2022})}\BibitemShut {NoStop}%
\bibitem [{\citenamefont {Harunari}\ \emph {et~al.}(2022)\citenamefont
  {Harunari}, \citenamefont {Dutta}, \citenamefont {Polettini},\ and\
  \citenamefont {Rold\'an}}]{Polettini}%
  \BibitemOpen
  \bibfield  {author} {\bibinfo {author} {\bibfnamefont {P.~E.}\ \bibnamefont
  {Harunari}}, \bibinfo {author} {\bibfnamefont {A.}~\bibnamefont {Dutta}},
  \bibinfo {author} {\bibfnamefont {M.}~\bibnamefont {Polettini}},\ and\
  \bibinfo {author} {\bibfnamefont {E.}~\bibnamefont {Rold\'an}},\ }\bibfield
  {title} {\bibinfo {title} {What to learn from a few visible transitions'
  statistics?},\ }\href {https://doi.org/10.1103/PhysRevX.12.041026} {\bibfield
   {journal} {\bibinfo  {journal} {Phys. Rev. X}\ }\textbf {\bibinfo {volume}
  {12}},\ \bibinfo {pages} {041026} (\bibinfo {year} {2022})}\BibitemShut
  {NoStop}%
\bibitem [{\citenamefont {Dieball}\ and\ \citenamefont {Godec}(2022)}]{Need}%
  \BibitemOpen
  \bibfield  {author} {\bibinfo {author} {\bibfnamefont {C.}~\bibnamefont
  {Dieball}}\ and\ \bibinfo {author} {\bibfnamefont {A.}~\bibnamefont
  {Godec}},\ }\bibfield  {title} {\bibinfo {title} {Mathematical,
  thermodynamical, and experimental necessity for coarse graining empirical
  densities and currents in continuous space},\ }\href
  {https://doi.org/10.1103/PhysRevLett.129.140601} {\bibfield  {journal}
  {\bibinfo  {journal} {Phys. Rev. Lett.}\ }\textbf {\bibinfo {volume} {129}},\
  \bibinfo {pages} {140601} (\bibinfo {year} {2022})}\BibitemShut {NoStop}%
\bibitem [{\citenamefont {van~der Meer}\ \emph {et~al.}(2023)\citenamefont
  {van~der Meer}, \citenamefont {Deg\"unther},\ and\ \citenamefont
  {Seifert}}]{Snippets}%
  \BibitemOpen
  \bibfield  {author} {\bibinfo {author} {\bibfnamefont {J.}~\bibnamefont
  {van~der Meer}}, \bibinfo {author} {\bibfnamefont {J.}~\bibnamefont
  {Deg\"unther}},\ and\ \bibinfo {author} {\bibfnamefont {U.}~\bibnamefont
  {Seifert}},\ }\bibfield  {title} {\bibinfo {title} {Time-resolved statistics
  of snippets as general framework for model-free entropy estimators},\ }\href
  {https://doi.org/10.1103/PhysRevLett.130.257101} {\bibfield  {journal}
  {\bibinfo  {journal} {Phys. Rev. Lett.}\ }\textbf {\bibinfo {volume} {130}},\
  \bibinfo {pages} {257101} (\bibinfo {year} {2023})}\BibitemShut {NoStop}%
\bibitem [{\citenamefont {Godec}\ and\ \citenamefont
  {Makarov}(2023{\natexlab{a}})}]{doi:10.1021/acs.jpclett.2c03244}%
  \BibitemOpen
  \bibfield  {author} {\bibinfo {author} {\bibfnamefont {A.}~\bibnamefont
  {Godec}}\ and\ \bibinfo {author} {\bibfnamefont {D.~E.}\ \bibnamefont
  {Makarov}},\ }\bibfield  {title} {\bibinfo {title} {Challenges in inferring
  the directionality of active molecular processes from single-molecule
  fluorescence resonance energy transfer trajectories},\ }\href
  {https://doi.org/10.1021/acs.jpclett.2c03244} {\bibfield  {journal} {\bibinfo
   {journal} {J. Phys. Chem. Lett.}\ }\textbf {\bibinfo {volume} {14}},\
  \bibinfo {pages} {49} (\bibinfo {year} {2023}{\natexlab{a}})}\BibitemShut
  {NoStop}%
\bibitem [{\citenamefont {Andrieux}(2012)}]{andrieux2012bounding}%
  \BibitemOpen
  \bibfield  {author} {\bibinfo {author} {\bibfnamefont {D.}~\bibnamefont
  {Andrieux}},\ }\bibfield  {title} {\bibinfo {title} {Bounding the coarse
  graining error in hidden markov dynamics},\ }\href
  {https://doi.org/10.1016/j.aml.2012.02.002} {\bibfield  {journal} {\bibinfo
  {journal} {Appl. Math. Lett.}\ }\textbf {\bibinfo {volume} {25}},\ \bibinfo
  {pages} {1734} (\bibinfo {year} {2012})}\BibitemShut {NoStop}%
\bibitem [{\citenamefont {Gomez-Marin}\ \emph {et~al.}(2008)\citenamefont
  {Gomez-Marin}, \citenamefont {Parrondo},\ and\ \citenamefont {Van~den
  Broeck}}]{PhysRevE.78.011107}%
  \BibitemOpen
  \bibfield  {author} {\bibinfo {author} {\bibfnamefont {A.}~\bibnamefont
  {Gomez-Marin}}, \bibinfo {author} {\bibfnamefont {J.~M.~R.}\ \bibnamefont
  {Parrondo}},\ and\ \bibinfo {author} {\bibfnamefont {C.}~\bibnamefont
  {Van~den Broeck}},\ }\bibfield  {title} {\bibinfo {title} {Lower bounds on
  dissipation upon coarse graining},\ }\href
  {https://doi.org/10.1103/PhysRevE.78.011107} {\bibfield  {journal} {\bibinfo
  {journal} {Phys. Rev. E}\ }\textbf {\bibinfo {volume} {78}},\ \bibinfo
  {pages} {011107} (\bibinfo {year} {2008})}\BibitemShut {NoStop}%
\bibitem [{\citenamefont {Rahav}\ and\ \citenamefont
  {Jarzynski}(2007)}]{rahav2007fluctuation}%
  \BibitemOpen
  \bibfield  {author} {\bibinfo {author} {\bibfnamefont {S.}~\bibnamefont
  {Rahav}}\ and\ \bibinfo {author} {\bibfnamefont {C.}~\bibnamefont
  {Jarzynski}},\ }\bibfield  {title} {\bibinfo {title} {Fluctuation relations
  and coarse-graining},\ }\href
  {https://dx.doi.org/10.1088/1742-5468/2007/09/P09012} {\bibfield  {journal}
  {\bibinfo  {journal} {J. Stat. Mech.}\ }\textbf {\bibinfo {volume} {2007}},\
  \bibinfo {pages} {P09012} (\bibinfo {year} {2007})}\BibitemShut {NoStop}%
\bibitem [{\citenamefont {Teza}\ and\ \citenamefont
  {Stella}(2020)}]{PhysRevLett.125.110601}%
  \BibitemOpen
  \bibfield  {author} {\bibinfo {author} {\bibfnamefont {G.}~\bibnamefont
  {Teza}}\ and\ \bibinfo {author} {\bibfnamefont {A.~L.}\ \bibnamefont
  {Stella}},\ }\bibfield  {title} {\bibinfo {title} {Exact coarse graining
  preserves entropy production out of equilibrium},\ }\href
  {https://doi.org/10.1103/PhysRevLett.125.110601} {\bibfield  {journal}
  {\bibinfo  {journal} {Phys. Rev. Lett.}\ }\textbf {\bibinfo {volume} {125}},\
  \bibinfo {pages} {110601} (\bibinfo {year} {2020})}\BibitemShut {NoStop}%
\bibitem [{\citenamefont {Rold\'an}\ and\ \citenamefont
  {Parrondo}(2010)}]{PhysRevLett.105.150607}%
  \BibitemOpen
  \bibfield  {author} {\bibinfo {author} {\bibfnamefont {E.}~\bibnamefont
  {Rold\'an}}\ and\ \bibinfo {author} {\bibfnamefont {J.~M.~R.}\ \bibnamefont
  {Parrondo}},\ }\bibfield  {title} {\bibinfo {title} {Estimating dissipation
  from single stationary trajectories},\ }\href
  {https://doi.org/10.1103/PhysRevLett.105.150607} {\bibfield  {journal}
  {\bibinfo  {journal} {Phys. Rev. Lett.}\ }\textbf {\bibinfo {volume} {105}},\
  \bibinfo {pages} {150607} (\bibinfo {year} {2010})}\BibitemShut {NoStop}%
\bibitem [{\citenamefont {Rold\'an}\ and\ \citenamefont
  {Parrondo}(2012)}]{PhysRevE.85.031129}%
  \BibitemOpen
  \bibfield  {author} {\bibinfo {author} {\bibfnamefont {E.}~\bibnamefont
  {Rold\'an}}\ and\ \bibinfo {author} {\bibfnamefont {J.~M.~R.}\ \bibnamefont
  {Parrondo}},\ }\bibfield  {title} {\bibinfo {title} {Entropy production and
  kullback-leibler divergence between stationary trajectories of discrete
  systems},\ }\href {https://doi.org/10.1103/PhysRevE.85.031129} {\bibfield
  {journal} {\bibinfo  {journal} {Phys. Rev. E}\ }\textbf {\bibinfo {volume}
  {85}},\ \bibinfo {pages} {031129} (\bibinfo {year} {2012})}\BibitemShut
  {NoStop}%
\bibitem [{\citenamefont {Baiesi}\ \emph {et~al.}(2023)\citenamefont {Baiesi},
  \citenamefont {Falasco},\ and\ \citenamefont
  {Nishiyama}}]{baiesi2023effective}%
  \BibitemOpen
  \bibfield  {author} {\bibinfo {author} {\bibfnamefont {M.}~\bibnamefont
  {Baiesi}}, \bibinfo {author} {\bibfnamefont {G.}~\bibnamefont {Falasco}},\
  and\ \bibinfo {author} {\bibfnamefont {T.}~\bibnamefont {Nishiyama}},\
  }\href@noop {} {\bibinfo {title} {Effective estimation of entropy production
  with lacking data}} (\bibinfo {year} {2023}),\ \Eprint
  {https://arxiv.org/abs/2305.04657} {arXiv:2305.04657 [cond-mat.stat-mech]}
  \BibitemShut {NoStop}%
\bibitem [{\citenamefont {Berezhkovskii}\ and\ \citenamefont
  {Makarov}(2018)}]{berezhkovskii_single-molecule_2018}%
  \BibitemOpen
  \bibfield  {author} {\bibinfo {author} {\bibfnamefont {A.~M.}\ \bibnamefont
  {Berezhkovskii}}\ and\ \bibinfo {author} {\bibfnamefont {D.~E.}\ \bibnamefont
  {Makarov}},\ }\bibfield  {title} {\bibinfo {title} {Single-{Molecule} {Test}
  for {Markovianity} of the {Dynamics} along a {Reaction} {Coordinate}},\
  }\href {https://doi.org/10.1021/acs.jpclett.8b00956} {\bibfield  {journal}
  {\bibinfo  {journal} {J. Phys. Chem. Lett.}\ }\textbf {\bibinfo {volume}
  {9}},\ \bibinfo {pages} {2190} (\bibinfo {year} {2018})}\BibitemShut
  {NoStop}%
\bibitem [{\citenamefont {Engbring}\ \emph {et~al.}(2023)\citenamefont
  {Engbring}, \citenamefont {Boriskovsky}, \citenamefont {Roichman},\ and\
  \citenamefont {Lindner}}]{Lindner}%
  \BibitemOpen
  \bibfield  {author} {\bibinfo {author} {\bibfnamefont {K.}~\bibnamefont
  {Engbring}}, \bibinfo {author} {\bibfnamefont {D.}~\bibnamefont
  {Boriskovsky}}, \bibinfo {author} {\bibfnamefont {Y.}~\bibnamefont
  {Roichman}},\ and\ \bibinfo {author} {\bibfnamefont {B.}~\bibnamefont
  {Lindner}},\ }\bibfield  {title} {\bibinfo {title} {A nonlinear
  fluctuation-dissipation test for markovian systems},\ }\href
  {https://doi.org/10.1103/PhysRevX.13.021034} {\bibfield  {journal} {\bibinfo
  {journal} {Phys. Rev. X}\ }\textbf {\bibinfo {volume} {13}},\ \bibinfo
  {pages} {021034} (\bibinfo {year} {2023})}\BibitemShut {NoStop}%
\bibitem [{\citenamefont {Lapolla}\ and\ \citenamefont
  {Godec}(2021)}]{PhysRevResearch.3.L022018}%
  \BibitemOpen
  \bibfield  {author} {\bibinfo {author} {\bibfnamefont {A.}~\bibnamefont
  {Lapolla}}\ and\ \bibinfo {author} {\bibfnamefont {A.}~\bibnamefont
  {Godec}},\ }\bibfield  {title} {\bibinfo {title} {Toolbox for quantifying
  memory in dynamics along reaction coordinates},\ }\href
  {https://doi.org/10.1103/PhysRevResearch.3.L022018} {\bibfield  {journal}
  {\bibinfo  {journal} {Phys. Rev. Research}\ }\textbf {\bibinfo {volume}
  {3}},\ \bibinfo {pages} {L022018} (\bibinfo {year} {2021})}\BibitemShut
  {NoStop}%
\bibitem [{\citenamefont {Hartich}\ and\ \citenamefont {Godec}(2023)}]{Viol}%
  \BibitemOpen
  \bibfield  {author} {\bibinfo {author} {\bibfnamefont {D.}~\bibnamefont
  {Hartich}}\ and\ \bibinfo {author} {\bibfnamefont {A.}~\bibnamefont
  {Godec}},\ }\bibfield  {title} {\bibinfo {title} {Violation of local detailed
  balance upon lumping despite a clear timescale separation},\ }\href
  {https://doi.org/10.1103/PhysRevResearch.5.L032017} {\bibfield  {journal}
  {\bibinfo  {journal} {Phys. Rev. Res.}\ }\textbf {\bibinfo {volume} {5}},\
  \bibinfo {pages} {L032017} (\bibinfo {year} {2023})}\BibitemShut {NoStop}%
\bibitem [{\citenamefont {Godec}\ and\ \citenamefont
  {Makarov}(2023{\natexlab{b}})}]{God_Mak}%
  \BibitemOpen
  \bibfield  {author} {\bibinfo {author} {\bibfnamefont {A.}~\bibnamefont
  {Godec}}\ and\ \bibinfo {author} {\bibfnamefont {D.~E.}\ \bibnamefont
  {Makarov}},\ }\bibfield  {title} {\bibinfo {title} {Challenges in inferring
  the directionality of active molecular processes from single-molecule
  fluorescence resonance energy transfer trajectories},\ }\href
  {https://doi.org/10.1021/acs.jpclett.2c03244} {\bibfield  {journal} {\bibinfo
   {journal} {J. Phys. Chem. Lett.}\ }\textbf {\bibinfo {volume} {14}},\
  \bibinfo {pages} {49} (\bibinfo {year} {2023}{\natexlab{b}})}\BibitemShut
  {NoStop}%
\bibitem [{\citenamefont {Hartich}\ and\ \citenamefont
  {Godec}(2021{\natexlab{b}})}]{TUR_David}%
  \BibitemOpen
  \bibfield  {author} {\bibinfo {author} {\bibfnamefont {D.}~\bibnamefont
  {Hartich}}\ and\ \bibinfo {author} {\bibfnamefont {A.}~\bibnamefont
  {Godec}},\ }\bibfield  {title} {\bibinfo {title} {Thermodynamic uncertainty
  relation bounds the extent of anomalous diffusion},\ }\href
  {https://doi.org/10.1103/PhysRevLett.127.080601} {\bibfield  {journal}
  {\bibinfo  {journal} {Phys. Rev. Lett.}\ }\textbf {\bibinfo {volume} {127}},\
  \bibinfo {pages} {080601} (\bibinfo {year} {2021}{\natexlab{b}})}\BibitemShut
  {NoStop}%
\bibitem [{\citenamefont {Seifert}(2012)}]{seifert2012stochastic}%
  \BibitemOpen
  \bibfield  {author} {\bibinfo {author} {\bibfnamefont {U.}~\bibnamefont
  {Seifert}},\ }\bibfield  {title} {\bibinfo {title} {Stochastic
  thermodynamics, fluctuation theorems and molecular machines},\ }\href
  {https://iopscience.iop.org/article/10.1088/0034-4885/75/12/126001}
  {\bibfield  {journal} {\bibinfo  {journal} {Rep. Prog. Phys.}\ }\textbf
  {\bibinfo {volume} {75}},\ \bibinfo {pages} {126001} (\bibinfo {year}
  {2012})}\BibitemShut {NoStop}%
\bibitem [{\citenamefont {Monod}\ \emph {et~al.}(1965)\citenamefont {Monod},
  \citenamefont {Wyman},\ and\ \citenamefont {Changeux}}]{MWC}%
  \BibitemOpen
  \bibfield  {author} {\bibinfo {author} {\bibfnamefont {J.}~\bibnamefont
  {Monod}}, \bibinfo {author} {\bibfnamefont {J.}~\bibnamefont {Wyman}},\ and\
  \bibinfo {author} {\bibfnamefont {J.-P.}\ \bibnamefont {Changeux}},\
  }\bibfield  {title} {\bibinfo {title} {On the nature of allosteric
  transitions: A plausible model},\ }\href
  {https://doi.org/https://doi.org/10.1016/S0022-2836(65)80285-6} {\bibfield
  {journal} {\bibinfo  {journal} {J. Mol. Biol.}\ }\textbf {\bibinfo {volume}
  {12}},\ \bibinfo {pages} {88} (\bibinfo {year} {1965})}\BibitemShut {NoStop}%
\bibitem [{\citenamefont {Tu}(2008)}]{Tu}%
  \BibitemOpen
  \bibfield  {author} {\bibinfo {author} {\bibfnamefont {Y.}~\bibnamefont
  {Tu}},\ }\bibfield  {title} {\bibinfo {title} {The nonequilibrium mechanism
  for ultrasensitivity in a biological switch: Sensing by maxwell’s demons},\
  }\href {https://doi.org/10.1073/pnas.0804641105} {\bibfield  {journal}
  {\bibinfo  {journal} {Proc. Natl. Acad. Sci.}\ }\textbf {\bibinfo {volume}
  {105}},\ \bibinfo {pages} {11737–11741} (\bibinfo {year}
  {2008})}\BibitemShut {NoStop}%
\bibitem [{\citenamefont {Korobkova}\ \emph {et~al.}(2006)\citenamefont
  {Korobkova}, \citenamefont {Emonet}, \citenamefont {Park},\ and\
  \citenamefont {Cluzel}}]{Cluzel}%
  \BibitemOpen
  \bibfield  {author} {\bibinfo {author} {\bibfnamefont {E.~A.}\ \bibnamefont
  {Korobkova}}, \bibinfo {author} {\bibfnamefont {T.}~\bibnamefont {Emonet}},
  \bibinfo {author} {\bibfnamefont {H.}~\bibnamefont {Park}},\ and\ \bibinfo
  {author} {\bibfnamefont {P.}~\bibnamefont {Cluzel}},\ }\bibfield  {title}
  {\bibinfo {title} {Hidden stochastic nature of a single bacterial motor},\
  }\href {https://doi.org/10.1103/PhysRevLett.96.058105} {\bibfield  {journal}
  {\bibinfo  {journal} {Phys. Rev. Lett.}\ }\textbf {\bibinfo {volume} {96}},\
  \bibinfo {pages} {058105} (\bibinfo {year} {2006})}\BibitemShut {NoStop}%
\bibitem [{\citenamefont {Marzen}\ \emph {et~al.}(2013)\citenamefont {Marzen},
  \citenamefont {Garcia},\ and\ \citenamefont
  {Phillips}}]{marzen2013statistical}%
  \BibitemOpen
  \bibfield  {author} {\bibinfo {author} {\bibfnamefont {S.}~\bibnamefont
  {Marzen}}, \bibinfo {author} {\bibfnamefont {H.~G.}\ \bibnamefont {Garcia}},\
  and\ \bibinfo {author} {\bibfnamefont {R.}~\bibnamefont {Phillips}},\
  }\bibfield  {title} {\bibinfo {title} {Statistical mechanics of
  monod--wyman--changeux (mwc) models},\ }\href
  {https://doi.org/10.1016/j.jmb.2013.03.013} {\bibfield  {journal} {\bibinfo
  {journal} {J. Mol. Biol.}\ }\textbf {\bibinfo {volume} {425}},\ \bibinfo
  {pages} {1433} (\bibinfo {year} {2013})}\BibitemShut {NoStop}%
\bibitem [{\citenamefont {Berlaga}\ and\ \citenamefont
  {Kolomeisky}(2021)}]{Tolya_1}%
  \BibitemOpen
  \bibfield  {author} {\bibinfo {author} {\bibfnamefont {A.}~\bibnamefont
  {Berlaga}}\ and\ \bibinfo {author} {\bibfnamefont {A.~B.}\ \bibnamefont
  {Kolomeisky}},\ }\bibfield  {title} {\bibinfo {title} {Molecular mechanisms
  of active transport in antiporters: Kinetic constraints and efficiency},\
  }\href {https://doi.org/10.1021/acs.jpclett.1c02846} {\bibfield  {journal}
  {\bibinfo  {journal} {J. Phys. Chem. Lett.}\ }\textbf {\bibinfo {volume}
  {12}},\ \bibinfo {pages} {9588–9594} (\bibinfo {year} {2021})}\BibitemShut
  {NoStop}%
\bibitem [{\citenamefont {Berlaga}\ and\ \citenamefont
  {Kolomeisky}(2022{\natexlab{a}})}]{Tolya_2}%
  \BibitemOpen
  \bibfield  {author} {\bibinfo {author} {\bibfnamefont {A.}~\bibnamefont
  {Berlaga}}\ and\ \bibinfo {author} {\bibfnamefont {A.~B.}\ \bibnamefont
  {Kolomeisky}},\ }\bibfield  {title} {\bibinfo {title} {Understanding
  mechanisms of secondary active transport by analyzing the effects of
  mutations and stoichiometry},\ }\href
  {https://doi.org/10.1021/acs.jpclett.2c01232} {\bibfield  {journal} {\bibinfo
   {journal} {J. Phys. Chem. Lett.}\ }\textbf {\bibinfo {volume} {13}},\
  \bibinfo {pages} {5405–5412} (\bibinfo {year}
  {2022}{\natexlab{a}})}\BibitemShut {NoStop}%
\bibitem [{\citenamefont {Berlaga}\ and\ \citenamefont
  {Kolomeisky}(2022{\natexlab{b}})}]{Tolya_3}%
  \BibitemOpen
  \bibfield  {author} {\bibinfo {author} {\bibfnamefont {A.}~\bibnamefont
  {Berlaga}}\ and\ \bibinfo {author} {\bibfnamefont {A.~B.}\ \bibnamefont
  {Kolomeisky}},\ }\bibfield  {title} {\bibinfo {title} {Theoretical study of
  active secondary transport: Unexpected differences in molecular mechanisms
  for antiporters and symporters},\ }\bibfield  {journal} {\bibinfo  {journal}
  {J. Chem. Phys.}\ }\textbf {\bibinfo {volume} {156}},\ \href
  {https://doi.org/10.1063/5.0082589} {10.1063/5.0082589} (\bibinfo {year}
  {2022}{\natexlab{b}})\BibitemShut {NoStop}%
\bibitem [{\citenamefont {Hartich}\ \emph {et~al.}(2015)\citenamefont
  {Hartich}, \citenamefont {Barato},\ and\ \citenamefont
  {Seifert}}]{hartich2015nonequilibrium}%
  \BibitemOpen
  \bibfield  {author} {\bibinfo {author} {\bibfnamefont {D.}~\bibnamefont
  {Hartich}}, \bibinfo {author} {\bibfnamefont {A.~C.}\ \bibnamefont
  {Barato}},\ and\ \bibinfo {author} {\bibfnamefont {U.}~\bibnamefont
  {Seifert}},\ }\bibfield  {title} {\bibinfo {title} {Nonequilibrium sensing
  and its analogy to kinetic proofreading},\ }\href
  {http://dx.doi.org/10.1088/1367-2630/17/5/055026} {\bibfield  {journal}
  {\bibinfo  {journal} {New J. Phys.}\ }\textbf {\bibinfo {volume} {17}},\
  \bibinfo {pages} {055026} (\bibinfo {year} {2015})}\BibitemShut {NoStop}%
\bibitem [{\citenamefont {Owen}\ \emph {et~al.}(2020)\citenamefont {Owen},
  \citenamefont {Gingrich},\ and\ \citenamefont
  {Horowitz}}]{PhysRevX.10.011066}%
  \BibitemOpen
  \bibfield  {author} {\bibinfo {author} {\bibfnamefont {J.~A.}\ \bibnamefont
  {Owen}}, \bibinfo {author} {\bibfnamefont {T.~R.}\ \bibnamefont {Gingrich}},\
  and\ \bibinfo {author} {\bibfnamefont {J.~M.}\ \bibnamefont {Horowitz}},\
  }\bibfield  {title} {\bibinfo {title} {Universal thermodynamic bounds on
  nonequilibrium response with biochemical applications},\ }\href
  {https://doi.org/10.1103/PhysRevX.10.011066} {\bibfield  {journal} {\bibinfo
  {journal} {Phys. Rev. X}\ }\textbf {\bibinfo {volume} {10}},\ \bibinfo
  {pages} {011066} (\bibinfo {year} {2020})}\BibitemShut {NoStop}%
\bibitem [{\citenamefont {Lapolla}\ \emph {et~al.}(2021)\citenamefont
  {Lapolla}, \citenamefont {Smith},\ and\ \citenamefont
  {Godec}}]{lapolla2021ubiquitous}%
  \BibitemOpen
  \bibfield  {author} {\bibinfo {author} {\bibfnamefont {A.}~\bibnamefont
  {Lapolla}}, \bibinfo {author} {\bibfnamefont {J.~C.}\ \bibnamefont {Smith}},\
  and\ \bibinfo {author} {\bibfnamefont {A.}~\bibnamefont {Godec}},\
  }\href@noop {} {\bibinfo {title} {Ubiquitous dynamical time asymmetry in
  measurements on materials and biological systems}} (\bibinfo {year} {2021}),\
  \Eprint {https://arxiv.org/abs/2102.01666} {arXiv:2102.01666
  [cond-mat.stat-mech]} \BibitemShut {NoStop}%
\end{thebibliography}%
\end{document}